\documentclass[preprint,aps,prc,showpacs,nofootinbib,secnumarabic,floatfix]{revtex4-1}

\usepackage{graphicx}
\usepackage{bm}
\usepackage{color}
\usepackage{gensymb}
\usepackage{amsmath}
\usepackage{slashed}
\usepackage{appendix}

\newcommand{\ltsim}{\protect\raisebox{-0.5ex}{$\:\stackrel{\textstyle <}
	{\sim}\:$}}

\newcommand{\bvec}[1]{\ensuremath{\boldsymbol{#1}}}

\begin{document}

\title{Color transparency in $\bar p d \to \pi^- \pi^0 p$ reaction}  

\author{A.B. Larionov$^{1,2}$\footnote{Corresponding author.\\ 
        E-mail address: larionov@fias.uni-frankfurt.de},
        M. Strikman$^3$}

\affiliation{$^1$Institut f\"ur Kernphysik, Forschungszentrum J\"ulich, D-52425 J\"ulich, Germany\\
             $^2$Frankfurt Institute for Advanced Studies (FIAS), 
             D-60438 Frankfurt am Main, Germany\\
             $^3$Pennsylvania State University, University Park, PA 16802, USA}

\begin{abstract}
  We consider exclusive two-pion production in antiproton-deuteron interactions at the beam
  momenta around 10 GeV/c in the kinematics with large momentum transfer in the underlying hard process
  $\bar p n \to \pi^- \pi^0$. The calculations are performed taking into account the antiproton and pion
  soft rescattering on the spectator proton in the framework of the generalized eikonal approximation.
  We focus on the color transparency effect that is modeled by introducing the dependence of rescattering amplitudes
  on the relative position of the struck and spectator nucleons along the momentum of a fast particle.
  As a consequence of the interplay between the impulse approximation and rescattering amplitudes the nuclear transparency ratio
  reveals a pretty complicated behaviour as a function of the transverse momentum of the spectator proton
  and the relative azimuthal angle between the $\pi^-$-meson and the proton.
  Color transparency significantly suppresses rescattering amplitudes which leads to substantial
  modifications of the nuclear transparency ratio moving it closer to the value obtained in the impulse approximation.
  By performing the Monte-Carlo analysis we determine that this effect can be studied at PANDA with a reasonable
  statistics.
\end{abstract}

\maketitle

\section{Introduction}
\label{intro}

It is widely accepted starting with the classical paper of F. Low \cite{Low:1975sv} that small size color singlets interact with hadrons with  a reduced strength
which, however, grows with energy since the cross section is proportional to the gluon strength at small $x$ \cite{Frankfurt:1993it,Blaettel:1993rd,Frankfurt:1996ri}.
This pattern is a distinctive feature of QCD. The first clear evidence of the small size configurations in hadrons was found in coherent diffractive dissociation
of 500 GeV/c pion into dijets on carbon and platinum targets \cite{Aitala:2000hc}. The pattern of the centrality dependence of the forward (large $x$)
jet production in pA scattering at LHC \cite{ATLAS:2014cpa,Chatrchyan:2014hqa} and RHIC \cite{Adare:2015gla} is naturally explained 
by the dominance of small size quark configurations containing  large $x$ \cite{Alvioli:2017wou}.

The small size -- or so-called point - like configurations (PLCs), are dynamically selected in the processes with large momentum transfer, $Q^2 \gg 1$ GeV$^2$
\cite{Brodsky,Mueller,Brodsky:1988xz}.
To probe this property of QCD, sufficiently high energies are necessary so that a hadron remains frozen in PLC over
distances comparable with the mean free path of the hadron in medium.
As a result, in the limit of high energies, the cross section of the processes with large $Q^2$
is proportional to the number of nucleons in the target -- the regime usually referred to as color transparency (CT),
see ref. \cite{Dutta:2012ii} for the most recent review.

In the case of exclusive processes $\gamma^*_L + N \to \mbox{meson + Baryon}$
induced by a longitudinally polarized virtual photon\footnote{In contrast to the real photon, the virtual one
is allowed to be longitudinal, i.e. to have helicity equal to zero.} $\gamma^*_L$
in the limit $x = \mbox{const}, Q^2 \to +\infty$
the factorization theorem \cite{Collins:1996fb} is valid which leads to the CT regime. The reason for CT is that the
longitudinally polarized photon consists of a $q\bar q$ pair having a transverse size $\sim 1/Q$.

There are  several classes of the exclusive  processes where large momentum transfer 
may lead to the transition to the CT regime. In particular, the increase of the nuclear
transparency with $Q^2$ was observed in virtual photon - nucleus interactions 
$\gamma^* A \to \pi^+ + A^*$ \cite{Clasie:2007aa} and $\gamma^* A \to \rho^0 + A^*$ \cite{ElFassi:2012nr}
studied at TJNAF. However, no clear evidence of CT was reported so far for processes with a
free nucleon in the final state.

For the hadron-induced semi-exclusive reactions $h+A \to h +p + (A-1)^*$ with large momentum transfer 
CT has been predicted in refs. \cite{Brodsky,Mueller}. Until now, only proton-induced processes $A(p,pp)(A-1)^*$
with the elastic scattering $pp \to pp$ at $\Theta_{c.m.}=90\degree$ from that reaction family have been studied
experimentally indicating the rise of nuclear transparency with beam momentum at $p_{\rm lab} < 9$ GeV/c consistent with CT
\cite{Carroll:1988rp,Mardor:1998zf,Leksanov:2001ui,Aclander:2004zm}. However, at higher $p_{\rm lab}$ the decrease
of nuclear transparency has been observed. It was suggested that this pattern could be explained by the interference
of hard perturbative and soft non-perturbative processes \cite{Brodsky:1987xw,Ralston:1988rb,VanOvermeire:2006tk}. 

In the case of antiproton beam, a large number of annihilation channels open which makes possible the
CT studies for reactions $A(\bar p, M_1M_2)$ with various mesonic final states $M=\pi, \eta, K, \rho \ldots$.
Since a $q \bar q$-pair is expected to be in a PLC with much larger probability
than a $qqq$-triple for the same $Q^2$, more pronounced CT signal is expected for mesons than for baryons.

In general, the  presence of CT is a necessary condition for the applicability of the factorization in the description
of the underlying hard elementary process, since without CT the multiple soft gluon exchanges prior and after the hard process
would not be suppressed.

The observation of CT for the two-meson final states is useful for the isolation of the type of the dominating pQCD diagrams. If the elementary
$\bar p N \to M_1M_2$ annihilation amplitude was dominated by the 
minimum number of the exchanged gluons in the limit $s, -t, -u \to +\infty, t/s={\rm const}$, then all
propagators would be highly virtual ($\propto 1/s$) leading to the scaling law of refs. \cite{Brodsky:1973kr,Matveev:1973ra}.
High virtuality of the propagators in momentum space implies that all interaction vertices are near each other in configuration space
(see, for example, ref. \cite{Carlson:1992jw}).
This essentially corresponds to the formation of PLCs unavoidably leading to CT.
On the other hand, the dominating disconnected graphs are likely to correspond to the normal hadronic size of the
participating quark configurations leading to the 'normal' Glauber-like nuclear transparency.
The interference of the connected and disconnected diagrams may result in a complicated oscillation pattern of the nuclear transparency,
similar to the case of $A(p,pp)$ reaction \cite{Ralston:1988rb}.

In this work, we will consider the reaction $d(\bar p,\pi^- \pi^0)p$ at large momentum transfer in the elementary $\bar p n \to \pi^- \pi^0$
process. We will focus on the beam momentum region of 5-15 GeV/c which could be studied by the PANDA experiment at FAIR \cite{PANDA}.
One motivation of our study are the strong CT effects predicted in ref. \cite{Frankfurt:1996uz} for the $d(p,2p)n$ reaction at
the similar kinematical conditions.
So far the $d(p,2p)n$ reaction at large momentum transfer has not been studied experimentally. 

The observation of CT at such relatively low beam momenta is, however, difficult since PLCs are not stable.
They expand to the normal hadronic size on the length scale given by the coherence length (see Eq. (\ref{l_h}) below).
At momenta $\sim 10$ GeV/c, the coherence length ($\sim 4-6$ fm) is comparable to the radii of medium size nuclei.
Thus, the PLC transforms to a normally interacting hadron already within the nuclear interior. This reduces the increase
of the nuclear transparency which is the main signal of CT.

The deuteron target -- in spite of the weakness of the initial and final state interactions -- gives
a unique opportunity to minimize the expansion of PLCs because the rescattering in the deuteron takes place on a relatively short distances
$r \leq 2$ fm estimated from the maxima of the $^3S_1$ ($r=1.7$ fm) and  $^3D_1$ ($r=1.4$ fm) components of the deuteron wave function (DWF)
\footnote{Throughout this work we apply the Paris potential model \cite{Lacombe:1981eg} for the DWF.}
This is somewhat conservative estimate as rescattering selects configurations smaller than average.
Hence, in the kinematics sensitive to rescattering CT should have a big effect.

The paper is organized as follows.  
In the next section we explain the underlying model based on the generalized eikonal approximation (GEA).
The model includes a single-step amplitude and three two-step amplitudes with elastic rescattering of the incoming antiproton and either
of the two outgoing pions on the spectator proton. The CT effects are included in the framework of the quantum diffusion model (QDM) via introducing
the dependence of elastic rescattering amplitude on the relative position of the proton and neutron projected on the particle momentum.
Sec. \ref{results} contains the results of the  numerical calculations of f
differential cross sections and 
transparency ratio. We show that the CT effects strongly modify the shape and the value of the transparency ratio as a function of the transverse momentum of the spectator and
of the azimuthal angle between one the pions and the spectator. In sec. \ref{MC} we provide the estimates of event rates at PANDA and
demonstrate some selected results of the Monte-Carlo simulations. The summary of the results, conclusions and directions for further
studies are presented in sec. \ref{summary}. The elementary amplitudes are described in Appendix \ref{ElemAmpl}.  

\section{The model}
\label{model}

At the beam momenta of $\sim 5-15$ GeV/c the differential cross sections of charge exchange $d\sigma_{\bar p p \to \bar n n}/dt$
\cite{Lee:1973zu}, $d\sigma_{\pi^- p \to \pi^0 n}/dt$ \cite{Apel:1977wy}
and the inelastic cross section $d\sigma_{\pi^+ p \to \rho^+ p}/dt$  \cite{Evans:1973jh} at $t=0$ are about two orders of magnitude
smaller than the respective elastic differential cross sections at $t=0$. Hence the charge exchange processes and the transitions
$\rho \to \pi$ can be safely neglected and we will keep elastic rescattering only. 
\begin{figure}
  \includegraphics[scale = 0.60]{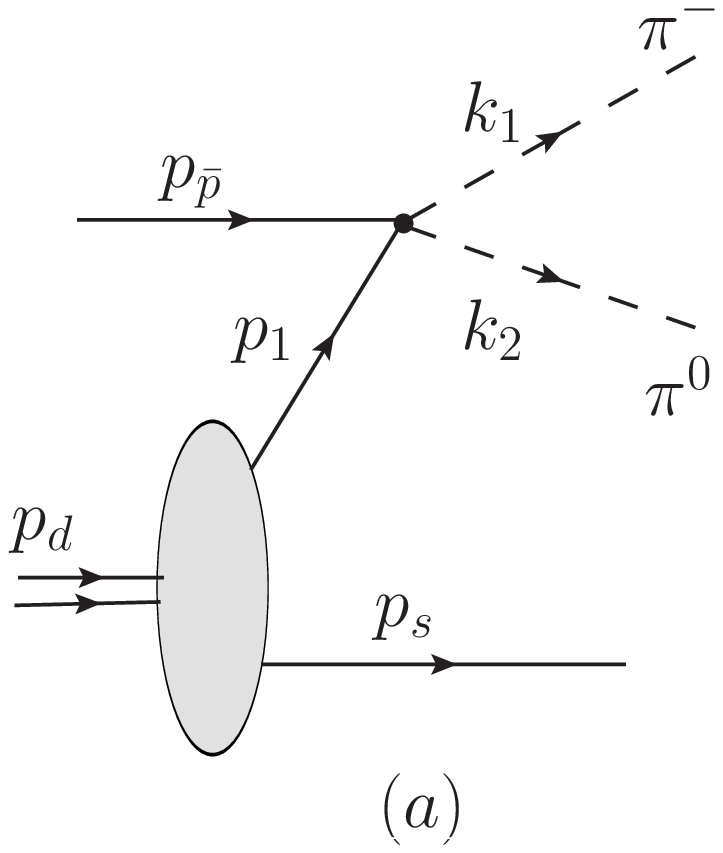}
  \includegraphics[scale = 0.60]{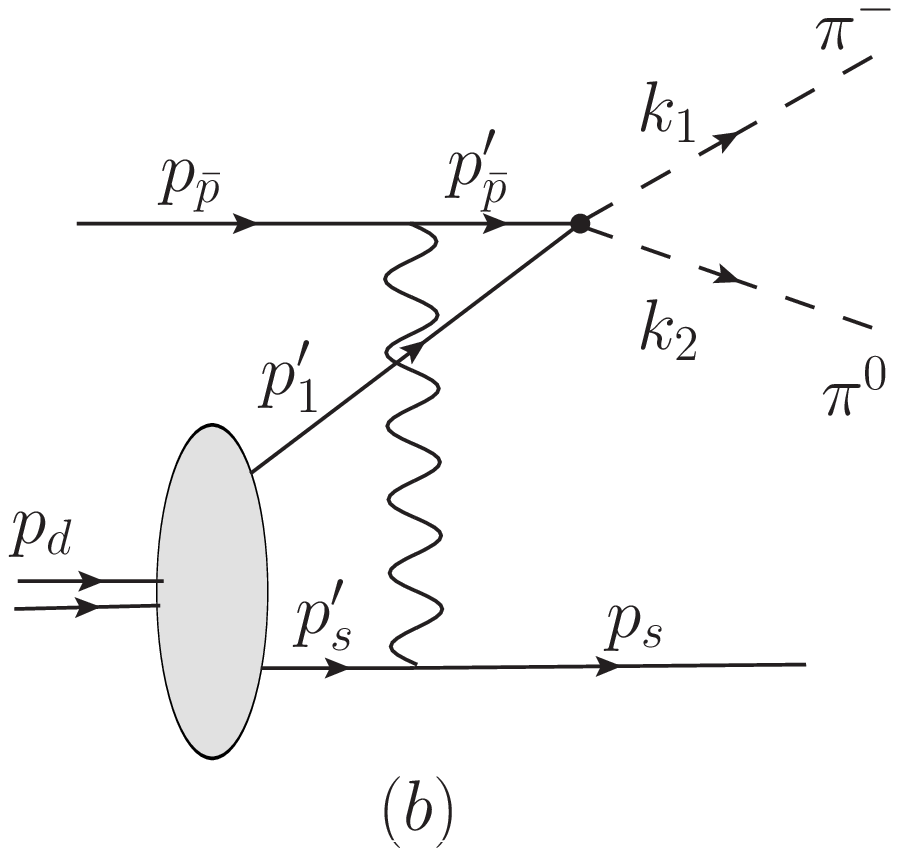}
  \includegraphics[scale = 0.60]{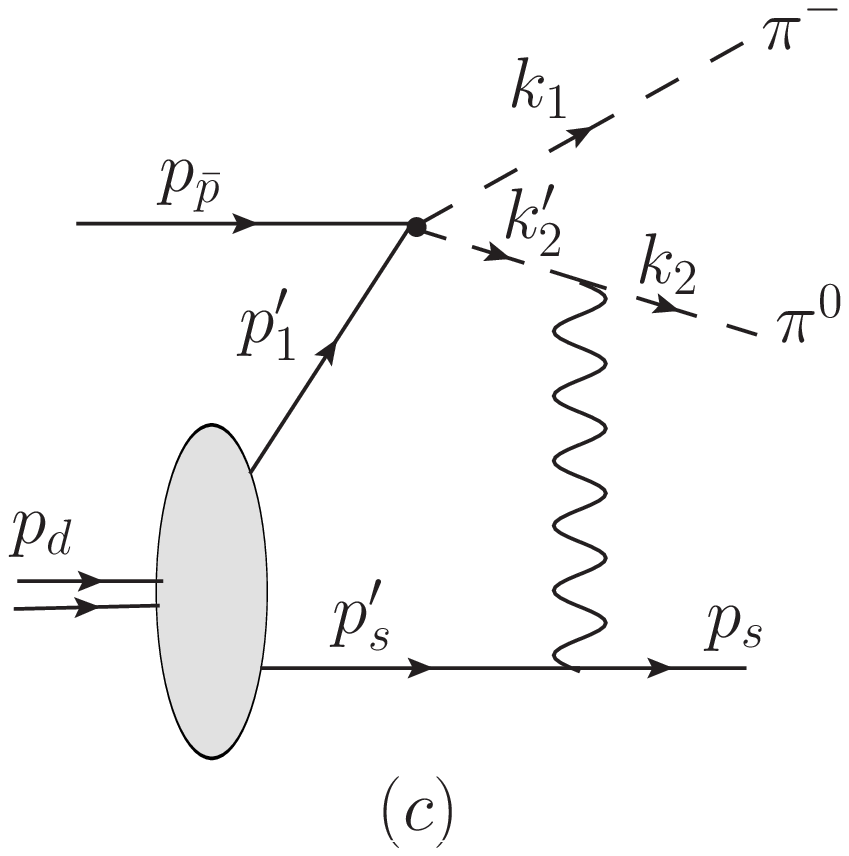}
  \includegraphics[scale = 0.60]{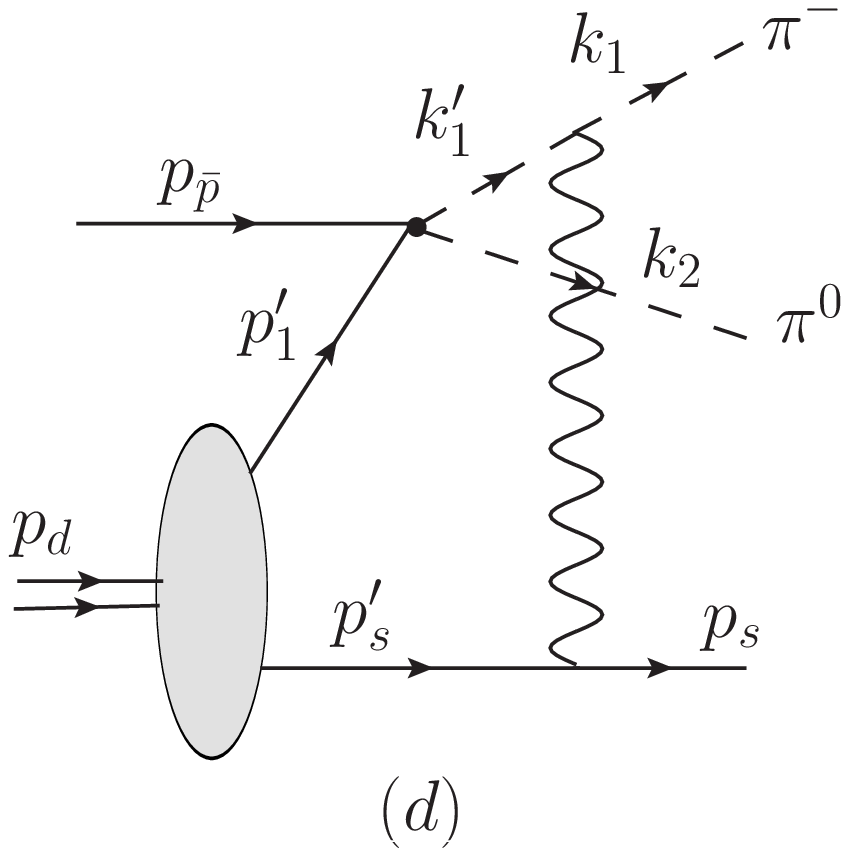}
\caption{\label{fig:diagr} Feynman diagrams for the process $\bar p d  \to \pi^- \pi^0 p$.
  The wavy lines denote soft elastic scattering amplitudes. The four momenta of the antiproton, deuteron, $\pi^-$, $\pi^0$
  and of the spectator proton are denoted as $p_{\bar p}$, $p_d$, $k_1$, $k_2$ and $p_s$, respectively. $p_1$ is the four momentum
  of the intermediate neutron in the IA amplitude (a). The primed quantities denote the four momenta of the corresponding
  intermediate particles in the amplitudes (b),(c),(d) with rescattering.}
\end{figure}
The contributions which we take into account are depicted in Fig.~\ref{fig:diagr}. The impulse approximation (IA) amplitude
\footnote{We always mean invariant amplitudes defined according to ref. \cite{BLP}.} 
(a) is expressed as
\begin{equation}
    M^{(a)} = M_{\rm ann}(k_1,k_2,p_{\bar p}) 
              \frac{i\Gamma_{d \to pn}(p_d,p_s)}{p_1^2-m_N^2+i\epsilon}~,               \label{M^(a)}
\end{equation}
where $M_{\rm ann}(k_1,k_2,p_{\bar p})$ is the amplitude of the $\bar p n \to \pi^- \pi^0$ annihilation
and $\Gamma_{d \to pn}(p_d,p_s)$ is the $d \to p n$ vertex function. The sum over spin projection of the intermediate neutron
is implicitly assumed in Eq.(\ref{M^(a)}) and below. For the on-shell spectator proton the following relation holds
in the deuteron rest frame  in the non-relativistic description and
neglecting the non-nucleonic degrees of freedom in the deuteron (cf. \cite{Larionov:2018lpk}):
\begin{equation} 
 \frac{i\Gamma_{d \to pn}(p_d,p_s)}{p_1^2-m_N^2+i\epsilon} 
 = \left(\frac{2E_sm_d}{p_1^0}\right)^{1/2} (2\pi)^{3/2} \phi(\bvec{p}_1)~,~~\bvec{p}_1=-\bvec{p}_s~,     \label{Gamma_d}
\end{equation}
where $\phi(\bvec{p}_1)$ is the momentum space DWF which contains both S- and D-waves normalized as
\begin{equation}
   \int d^3p |\phi(\bvec{p})|^2 = 1~,    \label{DWFnorm}
\end{equation}
$E_s=\sqrt{m_N^2+\bvec{p}_s^2}$ and $p_1^0=m_d-E_s$ are the energies of the on-shell spectator proton and of the
off-shell struck neutron, respectively.

The amplitudes with rescattering of the proton (b), $\pi^0$ (c), and $\pi^-$ (d) are given by the following expressions:
\begin{eqnarray}
  M^{(b)} &=& \int \frac{d^4p_s^\prime}{(2\pi)^4}
             \frac{M_{\bar p p}(p_s,p_s^\prime,p_{\bar p}) M_{\rm ann}(k_1,k_2,p_{\bar p}^\prime)
             \Gamma_{d \to pn}(p_d,p_s^\prime)}%
             {(p_{\bar p}^{\prime 2}-m_N^2+i\epsilon)(p_s^{\prime 2}-m_N^2+i\epsilon)(p_1^{\prime 2}-m_N^2+i\epsilon)}~,   \label{M^(b)} \\
  M^{(c)} &=& \int \frac{d^4p_s^\prime}{(2\pi)^4}
             \frac{M_{\pi^0 p}(k_2,p_s,p_s^\prime) M_{\rm ann}(k_1,k_2^\prime,p_{\bar p})
             \Gamma_{d \to pn}(p_d,p_s^\prime)}%
             {(k_2^{\prime 2}-m_\pi^2+i\epsilon)(p_s^{\prime 2}-m_N^2+i\epsilon)(p_1^{\prime 2}-m_N^2+i\epsilon)}~,   \label{M^(c)} \\
  M^{(d)} &=& \int \frac{d^4p_s^\prime}{(2\pi)^4}
             \frac{M_{\pi^- p}(k_1,p_s,p_s^\prime) M_{\rm ann}(k_1^\prime,k_2,p_{\bar p})
             \Gamma_{d \to pn}(p_d,p_s^\prime)}%
             {(k_1^{\prime 2}-m_\pi^2+i\epsilon)(p_s^{\prime 2}-m_N^2+i\epsilon)(p_1^{\prime 2}-m_N^2+i\epsilon)}~,   \label{M^(d)}
\end{eqnarray}
where $M_{\bar p p}(p_s,p_s^\prime,p_{\bar p})$, $M_{\pi^0 p}(k_2,p_s,p_s^\prime)$ and $M_{\pi^- p}(k_1,p_s,p_s^\prime)$
are the $\bar p p$, $\pi^0 p$, and $\pi^- p$ elastic scattering amplitudes, respectively.

We neglect the contribution of four double rescattering amplitudes considered in \cite{Frankfurt:1996uz} (Figs.~1e -- 1h).
Two of them are exactly zero (1e, 1f) while two others give a quite small contribution at the transverse momenta
of the spectator nucleon, $p_{st} \ltsim 400$ MeV/c, we consider in this paper.

Expressions (\ref{M^(b)}),(\ref{M^(c)}),(\ref{M^(d)})
follow from Feynman rules without additional assumptions. However, the singularities of the propagators 
do not allow for a direct numerical treatment.
Hence, for practical calculations one needs to make further simplifications depending on concrete kinematics.

At high energies one can neglect the dependence of the soft elastic rescattering amplitudes on the energy $p_s^{\prime 0}$
of the intermediate spectator. On the other hand, the hard annihilation amplitude varies weakly on the scale of the momenta transferred
in elastic rescattering and thus can be taken out of the integral. In these approximations, only the singularities due to the propagator
structure of Eqs.(\ref{M^(b)}),(\ref{M^(c)}),(\ref{M^(d)}) should be considered.

In the pion rescattering graphs (c,d), the only pole in the lower part of $p_s^{\prime 0}$ complex plane is
the particle pole of the spectator (see also ref. \cite{Frankfurt:1996uz}). Thus, Eqs.(\ref{M^(c)}),(\ref{M^(d)})
can be simplified by using the relation
\begin{equation}
    \int\limits_C \frac{dp_s^{\prime 0}}{2\pi} \frac{i}{p_s^{\prime 2} - m_N^2+ i\varepsilon}
   = \frac{1}{2E_s^\prime}~,~~
    E_s^\prime=\sqrt{\bvec{p}_s^{\prime 2}+m_N^2}~,    \label{p_s^prime_integr}
\end{equation}
with the contour $C$ closed in the lower part of the $p_s^{\prime 0}$ complex plane.
In the antiproton rescattering graph (b), it is convenient to perform the integration over $p_1^{\prime 0}=m_d-p_s^{\prime 0}$,
since in this case the particle pole of the struck neutron is the only pole in the lower part of the $p_1^{\prime 0}$
complex plane. This can be done by using Eqs.(\ref{p_s^prime_integr}) with replacement $s \to 1$.
After replacing $E_s^\prime, E_1^\prime \to m_N$,
i.e. neglecting Fermi motion in the calculation of nucleon energies,
we obtain the amplitudes (b),(c) and (d) in the pole approximation:
\begin{eqnarray}
  M^{(b)} &=& - m_N^{1/2} M_{\rm ann}(k_1,k_2,p_{\bar p}) \int \frac{d^3p_s^\prime}{(2\pi)^{3/2}} 
  \frac{\phi(\bvec{p}_1^\prime) M_{\bar p p}(t)}{p_{\bar p}^{\prime 2}-m_N^2+i\epsilon}~,   \label{M^(b)_pole} \\
  M^{(c)} &=& - m_N^{1/2} M_{\rm ann}(k_1,k_2,p_{\bar p}) \int \frac{d^3p_s^\prime}{(2\pi)^{3/2}} 
  \frac{\phi(\bvec{p}_1^\prime) M_{\pi^0 p}(t)}{k_2^{\prime 2}-m_\pi^2+i\epsilon}~,   \label{M^(c)_pole}\\
  M^{(d)} &=& - m_N^{1/2} M_{\rm ann}(k_1,k_2,p_{\bar p}) \int \frac{d^3p_s^\prime}{(2\pi)^{3/2}} 
  \frac{\phi(\bvec{p}_1^\prime) M_{\pi^- p}(t)}{k_1^{\prime 2}-m_\pi^2+i\epsilon}~,   \label{M^(d)_pole}       
\end{eqnarray}
where $\bvec{p}_1^\prime = -\bvec{p}_s^\prime$ and $t=k^2$ with $k=p_s-p_s^\prime$ being the four momentum
transfer to the spectator proton. 

The inverse propagator of an antiproton can be linearized 
with respect to the longitudinal momentum
transfer to the spectator:
\begin{equation}
p_{\bar p}^{\prime 2}-m_N^2+i\epsilon = (p_{\bar p}-k)^2-m_N^2+i\epsilon 
= -2p_{\bar p}k + k^2 +i\epsilon
=  2|\bvec{p}_{\bar p}| (k^z - \Delta^0_{\bar p} + i\epsilon)~, \label{pbar_inv_prop}
\end{equation}
with $z$ axis directed along $\bvec{p}_{\bar p}$ and
\begin{equation}
\Delta^0_{\bar p} \equiv |\bvec{p}_{\bar p}|^{-1}[E_{\bar p}(E_s-p_s^{\prime 0}) - (p_s- p_s^\prime)^2/2] 
     \simeq  |\bvec{p}_{\bar p}|^{-1} (E_{\bar p}+m_N)(E_s-m_N)~.  \label{Delta^0_pbar}
\end{equation}
In Eq.(\ref{Delta^0_pbar}) at the last step we used the relation $p_s^{\prime 0}=m_d-E_1^\prime$ with
$E_1^\prime=\sqrt{m_N^2+\bvec{p}_1^{\prime 2}}$ and then neglected the Fermi motion of the struck neutron
by setting $\bvec{p}_1^\prime=0$.

In a similar way, it is possible to transform the inverse propagator of the $\pi^0$-meson:
\begin{equation}
k_2^{\prime 2}-m_\pi^2+i\epsilon = (k_2+k)^2-m_\pi^2+i\epsilon = 2k_2k + k^2 +i\epsilon
= 2 |\bvec{k}_2| (-k^z + \Delta^0_2 +i\epsilon)~,  \label{pi^0_inv_prop}
\end{equation}
with  $z$ axis directed along $\bvec{k}_2$ and
\begin{equation}
  \Delta^0_2 \equiv |\bvec{k}_2|^{-1} [\omega_2(E_s-E_s^\prime) + (p_s-p_s^\prime)^2/2]
  \simeq  |\bvec{k}_2|^{-1} (\omega_2-m_N)(E_s-m_N)~,  \label{Delta^0_2}    
\end{equation}
with $\omega_2=\sqrt{m_\pi^2+\bvec{k}_2^2}$.
In the last step of Eq.(\ref{Delta^0_2}) the Fermi motion of the on shell intermediate spectator
is neglected, i.e. we set $\bvec{p}_s^\prime=0$.
The inverse propagator of the $\pi^-$-meson can be obtained by using Eqs.(\ref{pi^0_inv_prop}),(\ref{Delta^0_2})
with replacement of the subscript $2 \to 1$.

Eqs.(\ref{pbar_inv_prop}),(\ref{pi^0_inv_prop}) are the main assumptions of the GEA. Similar expressions appear
in other calculations based on the GEA (cf. \cite{Frankfurt:1994kt,Frankfurt:1996uz,Frankfurt:1996xx,Sargsian01,Larionov:2019mwa}).
We will proceed using the coordinate space representation which is necessary for the introduction of the CT effects later on.
The linearized propagators can be written down in the coordinate representation using the identity:
\begin{equation}
   \frac{i}{p+i\epsilon}
     =\int dz^0 \Theta(z^0) \mbox{e}^{i p z^0}~,   \label{Dcoord}
\end{equation}
where $\Theta(x)$ is the Heaviside step function ($\Theta(x)=0$ for $x<0$, $\Theta(x)=1/2$ for $x=0$, and $\Theta(x)=1$ for $x>0$).
The DWFs in the momentum and coordinate space are related as follows: 
\begin{equation}
   \phi(\bvec{p}_1^\prime) = \int \frac{d^3 r}{(2\pi)^{3/2}} \mbox{e}^{-i\bvec{p}_1^\prime\bvec{r}} 
                   \phi(\bvec{r})~, ~~~\bvec{r}=\bvec{r}_1-\bvec{r}_s~.     \label{phi(p_1^prime)}
\end{equation}

We will now put the intermediate fast particles on the mass shell by setting
$k^z = \Delta^0_{\bar p},\Delta^0_{2}$ and $\Delta^0_{1}$  in the elementary elastic rescattering amplitudes
$M_{\bar p p}(t)$, $M_{\pi^0 p}(t)$ and $M_{\pi^- p}(t)$, respectively.
This allows to perform integration over $p_s^{\prime z}$ in Eqs.(\ref{M^(b)_pole}),(\ref{M^(c)_pole}),(\ref{M^(d)_pole})
analytically. Then, after integration over azimuthal angle of $\bvec{k}_t$ we obtain the following expressions
for the amplitudes with elastic rescattering:
\begin{eqnarray}
  M^{(b)} &=& \frac{i M_{\rm ann}(k_1,k_2,p_{\bar p})}{4\pi |\bvec{p}_{\bar p}| m_N^{1/2}} \int d^3r \phi(\bvec{r})
  \Theta(z) 
  \mbox{e}^{i\bvec{p}_s\bvec{r}-i\Delta_{\bar p}^0z}
  \int\limits_0^{+\infty} dk_t  k_t M_{\bar p p}(t_{\bar p})    J_0(k_t b)~,        \label{M^(b)_coord}\\
  M^{(c)} &=& \frac{i M_{\rm ann}(k_1,k_2,p_{\bar p})}{4\pi |\bvec{k}_2| m_N^{1/2}} \int d^3\tilde{r} \phi(\bvec{r}) 
  \Theta(-\tilde{z}) 
  \mbox{e}^{i\bvec{p}_s\bvec{r}-i\Delta_{2}^0 \tilde{z}}
  \int\limits_0^{+\infty} dk_t  k_t M_{\pi^0 p}(t_2)  J_0(k_t \tilde{b})~,  \label{M^(c)_coord}\\
   M^{(d)} &=& \frac{i M_{\rm ann}(k_1,k_2,p_{\bar p})}{4\pi |\bvec{k}_1| m_N^{1/2}} \int d^3\tilde{r} \phi(\bvec{r}) 
  \Theta(-\tilde{z}) 
  \mbox{e}^{i\bvec{p}_s\bvec{r}-i\Delta_{1}^0 \tilde{z}}
  \int\limits_0^{+\infty} dk_t  k_t M_{\pi^- p}(t_1) J_0(k_t \tilde{b})~,        \label{M^(d)_coord}
\end{eqnarray}
where $t_j=(E_s-m_N)^2 - (\Delta^0_j)^2- k_t^2$ with $j=\bar p, 1, 2$, $b=(x^2+y^2)^{1/2}$, $\tilde{b}=(\tilde{x}^2+\tilde{y}^2)^{1/2}$,
and
\begin{equation}
  J_0(x)=\frac{1}{2\pi}\int\limits_0^{2\pi} d\phi\, \mbox{e}^{-ix\cos \phi}    \label{J_0}
\end{equation}
is the Bessel function of the first kind.
The space integration in the amplitude (\ref{M^(c)_coord}) is performed
in the rotated coordinate system with $\tilde{z}$-axis along $\bvec{k}_2=|\bvec{k}_2|(\sin\Theta \cos\phi, \sin\Theta \sin\phi, \cos\Theta)$.
This allows to accelerate numerical calculations, since the argument of the Bessel function does not depend on the azimuthal angle of $\tilde{\bvec{r}}$.
The position vector $\bvec{r}$ in the original frame with $z$-axis along the antiproton beam momentum is expressed via rotation matrix with
Euler angles $\alpha=\phi$, $\beta=\Theta$, and $\gamma=0$  \cite{VMKh}.
\begin{equation}
  \left(\begin{array}{l}
    x\\
    y\\
    z
  \end{array}
  \right)
  =\left( 
\begin{array}{lll}
      \cos\phi \cos\Theta & -\sin\phi & \cos\phi \sin\Theta \\
     \sin\phi \cos\Theta   &  \cos\phi & \sin\phi \sin\Theta \\
    -\sin\Theta                    &  0               & \cos\Theta
 \end{array}
\right)
  \left(\begin{array}{l}
    \tilde{x}\\
    \tilde{y}\\
    \tilde{z}
  \end{array}
  \right)~.         \label{rotation}
\end{equation} 
In a similar way, in the amplitude (\ref{M^(d)_coord}) the space integration is performed
in the rotated coordinate system with $\tilde{z}$-axis along $\bvec{k}_1$.
Note that the argument $\bvec{r}$
of the DWF was always calculated in the original frame with $z$-axis along the antiproton beam momentum that has been
fixed as the spin quantization axis.

For simplicity, the elastic rescattering amplitudes were supposed to conserve spin projections of particles
and to be spin-independent which is quite reasonable for small momentum transfers. The details of the elementary amplitudes
are given in Appendix \ref{ElemAmpl}. We will now discuss how the CT effects are included in the model discussed above. 

\subsection{The CT effects}
\label{CT}

CT implies that the hadrons participating in a hard collision are interacting with the surrounding nucleons in vicinity
of a hard collision point with a reduced strength. Since the transverse size of the hadrons is reduced, their form factors
in momentum space get harder which leads to the modification of the momentum transfer dependence of the elastic scattering.
These two effects are combined within the QDM \cite{Farrar:1988me,Frankfurt:1994kt} that is the model of CT at intermediate
energies where the expansion of PLCs is a non-negligible effect.
The QDM has been successfully applied to describe experimental data on the pion- \cite{Dutta:2003mk,Larson:2006ge,Cosyn:2007er}
and $\rho$-meson \cite{Frankfurt:2008pz,Gallmeister:2010wn} electroproduction at TJNAF.

In the QDM the elementary amplitudes of the elastic scattering become position dependent and can be written as follows:
\begin{equation}
    M_{h p}(t,z)= 2 i p_{h} m_N \sigma_{h p}^{\rm eff}(p_{h},|z|)
     (1-i\rho_{h p}) \mbox{e}^{B_{h p}t/2} 
     \frac{G_{h}(t \cdot \frac{\sigma_{h p}^{\rm eff}(p_{h},|z|)}{\sigma_{h p}^{\rm tot}})}{G_{h}(t)}~,  \label{M_hp_CT}
\end{equation}
where $z = (\bvec{r}_1-\bvec{r}_s) \cdot  \hat{\bvec{p}_{h}}$ is the relative position of the struck and spectator nucleons
along the hadron momentum $\bvec{p}_{h}$ ($h=\bar p, \pi$).
In comparison to the standard elastic scattering amplitudes of Eqs.(\ref{M_barpp}),(\ref{M_pipmp})
the amplitude of Eq.(\ref{M_hp_CT}) includes the effective cross section
\begin{equation}
   \sigma_{h p}^{\rm eff}(p_{h},|z|)
  = \sigma_{h p}^{\rm tot}\left(\left[ \frac{|z|}{l_{h}}
    + \frac{\langle n_{h}^2 k_{ht}^2\rangle}{Q^2} \left(1-\frac{|z|}{l_{h}}\right) \right]
    \Theta(l_{h}-|z|) +\Theta(|z|-l_{h})\right)~,                 \label{sigma_hp_eff}
\end{equation}
where $\sqrt{\langle k_{ht}^2\rangle}=0.35$ GeV/c is the average transverse momentum of a parton in the scattered hadron, 
$n_{h}$ is the number of valence partons in the hadron ($n_{\bar p}=3$, $n_{\pi}=2$).
$Q^2=\min(-t_{\rm hard},-u_{\rm hard})$ is the hard scale, $t_{\rm hard}=(p_{\bar p}-p_{\pi^-})^2,~~u_{\rm hard}=(p_{\bar p}-p_{\pi^0})^2$.
The effective cross section grows linearly with increasing longitudinal distance $|z|$ for $|z| \leq l_h$ and becomes equal
to the total hadron-proton cross section for $|z| \geq l_h$. Here the coherence length is expressed as 
\begin{equation}
  l_h = \frac{2p_{h}}{\Delta M^2}~,   \label{l_h}
\end{equation}
with the mass denominator $\Delta M^2 \simeq 0.7-1.1$ GeV$^2$ \cite{Frankfurt:1996uz,Dutta:2012ii}.

The reduced transverse size of the scattered hadron influences not only the total interaction strength with a proton, but
also the formfactor of this hadron which influences the momentum transfer dependence of the elastic amplitude \cite{Frankfurt:1994kt}.
This effect is taken into account by the last term in the r.h.s. of Eq.(\ref{M_hp_CT}) where $G_{h}(t)$ is the electromagnetic formfactor
of the scattered hadron. For the antiproton we apply the dipole formfactor
\begin{equation}
  G_{\bar p}(t)=\frac{1}{(1-t/0.71~\mbox{GeV}^2)^2}~,     \label{SachsFF}
\end{equation}  
that is the Sachs electric form factor of the proton. For the pion we use the monopole form factor
\begin{equation}
  G_\pi(t)=\frac{1}{1-  \langle r_\pi^2\rangle t/6}~,    \label{pionFF}
\end{equation}
where $\langle r_\pi^2\rangle = 0.439 \pm 0.008~\mbox{fm}^2$ is the mean square charge radius of the pion \cite{Amendolia:1986wj}.

\subsection{Observables}
\label{observables}

The differential cross section is defined as follows: 
\begin{equation}
  d\sigma_{\bar p d \to \pi_1^- \pi_2^0 p} =  
  \frac{(2\pi)^4\overline{|M|^2}}{4p_{\rm lab}m_d} d\Phi_3~,          \label{dsigma}
\end{equation}
where $\overline{|M|^2}=\overline{|M^{(a)}+M^{(b)}+M^{(c)}+M^{(d)}|^2}$ is the modulus squared of the total invariant amplitude
summed over spins of final particles and averaged over spins of initial particles and
\begin{equation}
  d\Phi_3 = \delta^{(4)}(p_{\bar p}+p_d-k_1-k_2-p_s)
  \frac{d^3k_1}{(2\pi)^32\omega_1}   \frac{d^3k_2}{(2\pi)^32\omega_2}   \frac{d^3p_s}{(2\pi)^32E_s}   \label{dPhi_3}
\end{equation}
is a Lorentz-invariant three-body phase space volume element. It is convenient to introduce the light
cone (LC) variables $\alpha_s$ and $\beta$. In the infinite momentum frame with fast backward deuteron, $\alpha_s/2$
is the deuteron momentum fraction carried by the spectator, while in the infinite momentum frame with fast forward antiproton,
$\beta/2$ is the fraction of the momentum of the $\bar p$+struck neutron system carried by $\pi^-$-meson.
\footnote{Of course, this has a meaning only if a frame exists that approximately satisfy the both conditions simultaneously.
  If we require highly-energetic $\bar p$, then such a frame could be chosen, for example, as the c.m. frame of the antiproton
  and the deuteron.}
Therefore, in the laboratory frame
with $\bar p$ beam momentum in positive $z$-direction
we have      
\begin{eqnarray}
  \alpha_s &=& \frac{2(E_s-p_s^z)}{m_d}~,                 \label{alpha_s}\\
  \beta    &=& \frac{2(\omega_1+k_1^z)}{E_{\bar p}+m_d-E_s+p_{\rm lab}-p_s^z}~.   \label{beta} 
\end{eqnarray}
Variables $\alpha_s$ and $\beta$ are longitudinal boost invariant and  approximately
conserved in elastic rescattering at high energies.
In the c.m. frame of the $\bar p$
and the struck neutron
one can express $\beta$ via the polar scattering angle
$\Theta_{c.m.}$ of the $\pi^-$-meson with respect to the 
antiproton:
\begin{equation}
  \beta \simeq 1 + \cos\Theta_{c.m.}~,     \label{beta_vs_Thetacm}
\end{equation}
where we neglected the pion mass and the change of $\beta$ due to a finite transverse momentum of the spectator.  
As shown in Appendix \ref{LCderiv}, by using the LC variables the four-differential cross section can be 
written in the following form : 
\begin{equation}
  \alpha_s \beta \frac{d^4\sigma}{d\alpha_s\, d\beta\, d\phi\, p_{s t} dp_{s t}}
  = \frac{\overline{|M|^2} k_{1t}}%
         {16(2\pi)^4 p_{\rm lab} m_d \kappa_t}~,                   \label{dsig/dalpha}
\end{equation}
where $\phi$ is the relative azimuthal angle between $\pi^-$-meson and spectator proton,
\begin{equation}
   \phi=\phi_1-\phi_s~,       \label{phi}
\end{equation}
and 
\begin{equation}
  \kappa_t = 2\left|\frac{2k_{1t}}{\beta} + p_{s t} \cos\phi \right|~.      \label{kappa_t}
\end{equation}

For a better visibility of the rescattering effects one can use the transparency ratio $T$. We will apply the definition
adopted from studies of $A(e,e^\prime p)$ and $d(p,2p)n$ reactions, see \cite{Frankfurt:1994kt,Frankfurt:1996uz} and refs. therein:
\begin{equation}
       T \equiv \frac{\sigma^{\rm DWIA}}{\sigma^{\rm IA}} 
      =  \frac{\overline{|M^{(a)}+M^{(b)}+M^{(c)}+M^{(d)}|^2}}{\overline{|M^{(a)}|^2}}~,      \label{T}
\end{equation}
where $\sigma^{\rm DWIA}$ and $\sigma^{\rm IA}$ are the differential cross sections calculated within the distorted wave impulse approximation
and impulse approximation, respectively.
Instead of $\sigma^{\rm DWIA}$  one should use a measured cross section if it is available.

We have to admit, however, that in this exploratory study we rely on a quite rough model for the amplitude of the $\bar p n \to \pi^- \pi^0$ channel. 
Thus, our calculations should be eventually normalized to the measured data for the $d(\bar p,\pi^- \pi^0)p$ process in the quasifree kinematics
$\alpha_s \simeq 1$,~~$p_{st}  \ltsim 0.1$ GeV/c, where the deviations from IA are small. 

\section{Results}
\label{results}

All calculations are performed in the transverse kinematics, $\alpha_s=1$, where the used non-relativistic description of the DWF
is expected to be valid (cf. \cite{Frankfurt:1996uz}).
\begin{figure}
  \includegraphics[scale = 0.50]{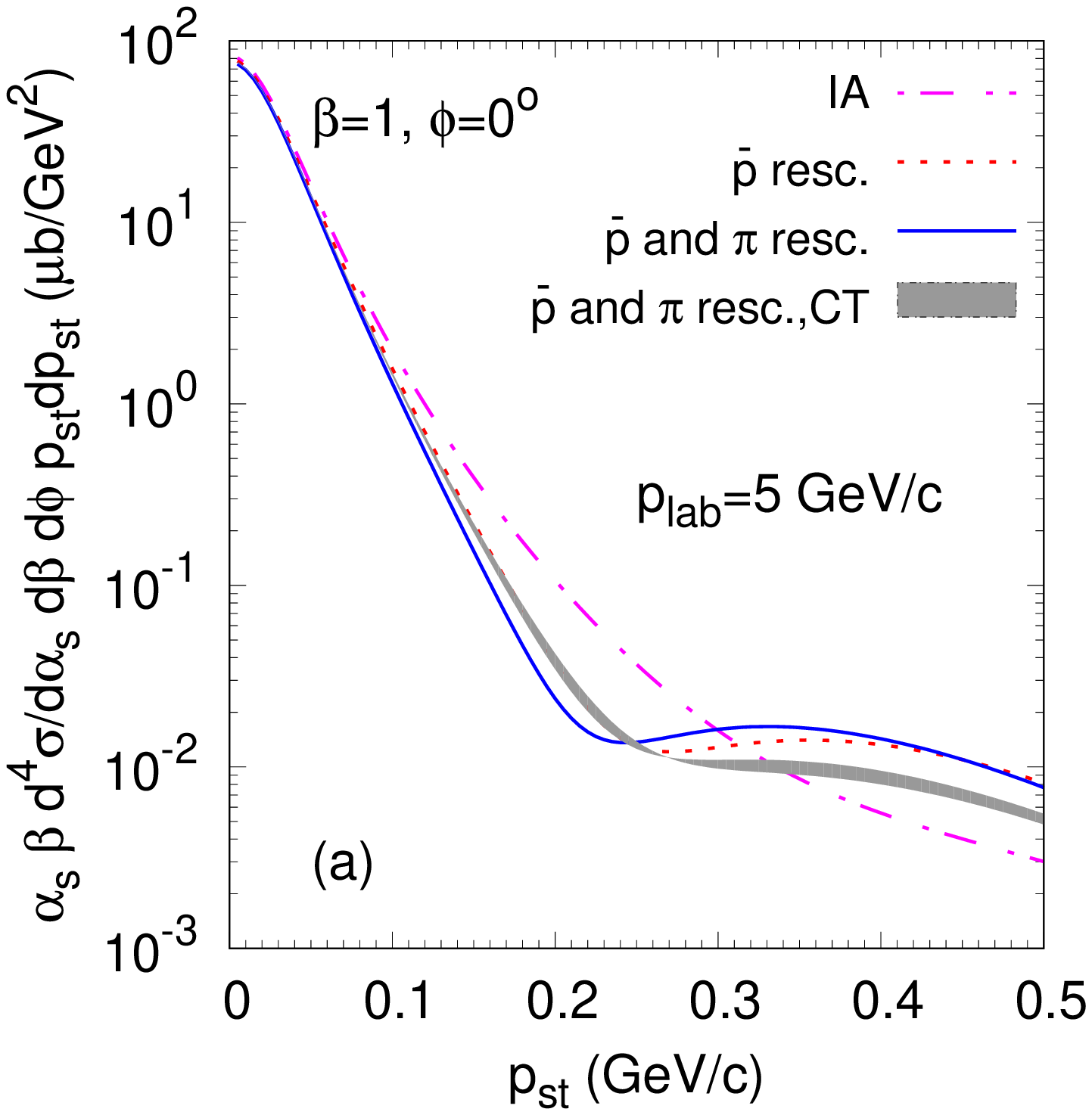}
  \includegraphics[scale = 0.50]{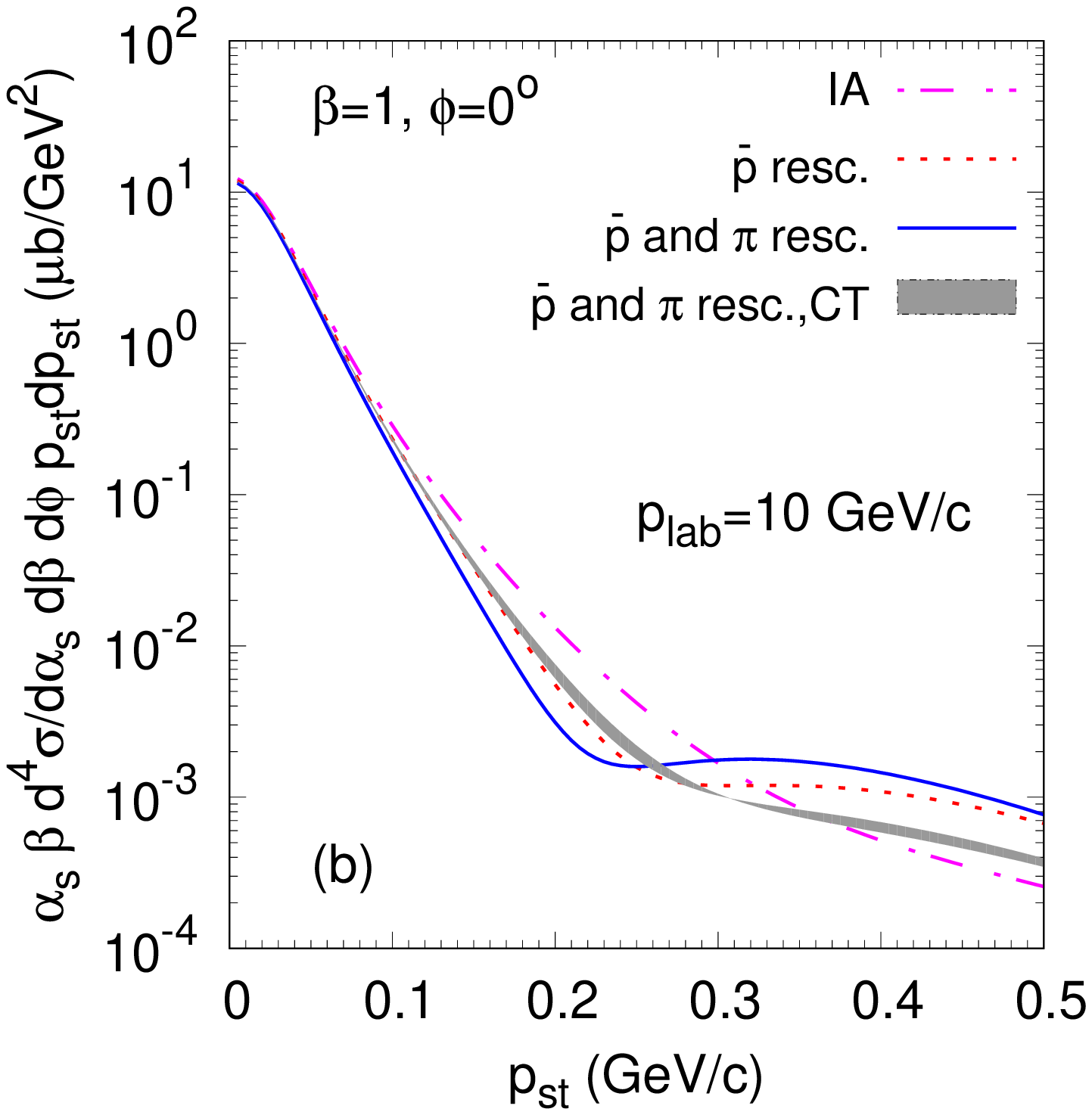}
  \includegraphics[scale = 0.50]{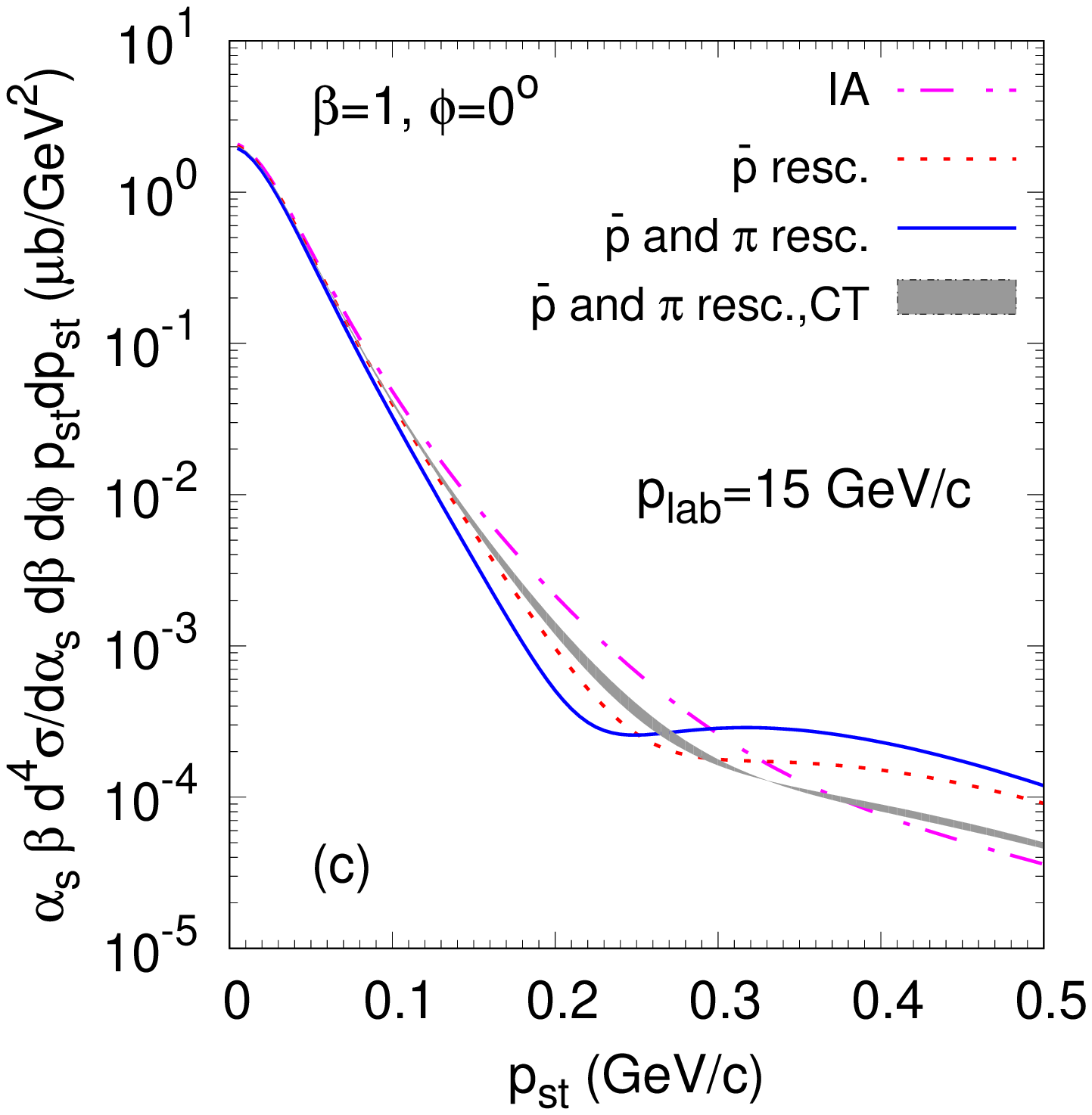}
  \includegraphics[scale = 0.50]{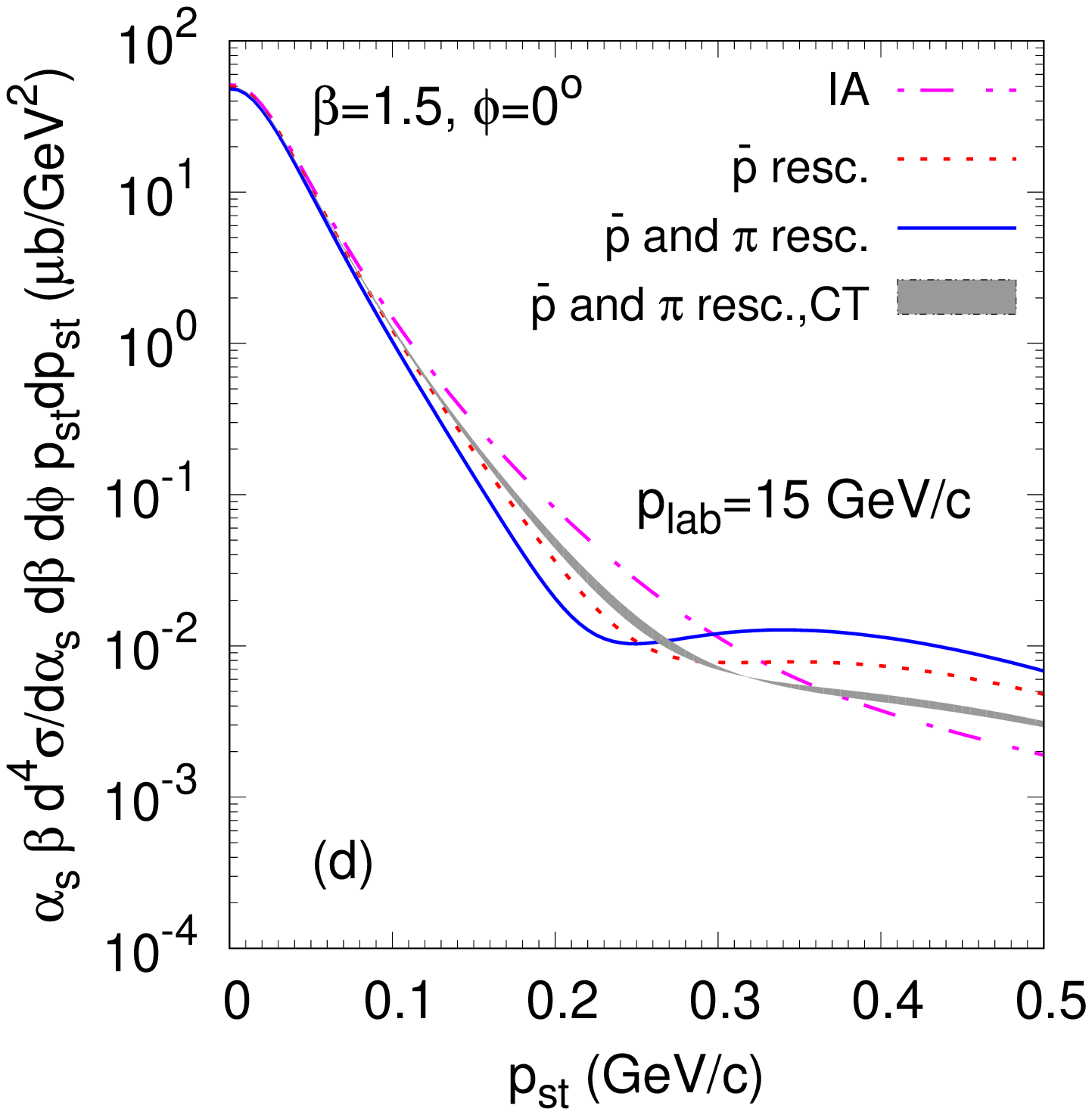}  
  \caption{\label{fig:sig_0deg} Four-differential cross section $\bar p d \to \pi^- \pi^0 p$ as a function of
    the transverse momentum of spectator proton for the relative azimuthal angle between $\pi^-$ and spectator proton
    $\phi=0\degree$.
    Different panels display the calculations for the different beam momenta $p_{\rm lab}$ and the LC momentum fraction
    $\beta$ carried by the $\pi^-$, as indicated. The GEA calculations are shown by the dash-dotted, dotted and
    solid lines corresponding to the IA, IA plus $\bar p$ rescattering, and IA plus $\bar p$ and pion rescattering.
    The calculations with $\bar p$ and pion rescattering taking into account CT are displayed by the grey band
    limited by the values of the mass denominator of the coherence length $\Delta M^2 = 0.7$ GeV$^2$ and 1.1 GeV$^2$.
    The calculations are performed for $\alpha_s=1$.}
\end{figure}
In Fig.~\ref{fig:sig_0deg} we display the spectator transverse momentum, $p_{st}$, dependence of the four-differential cross section,
Eq.(\ref{dsig/dalpha}), calculated for three values of the beam momentum, $p_{\rm lab}=5$, 10 and 15 GeV/c for the in-plane
kinematics, $\phi=0\degree$. Pure IA produces a monotonically dropping cross section with $p_{st}$.
Antiproton rescattering leads to strong deviations from IA: depletion at low and enhancement at high spectator transverse momenta.
Pion rescattering further amplifies these effects. CT diminishes the effects of rescattering driving the cross section closer
to the IA shape. The effect of CT is, as expected, stronger for larger beam momenta. However, due to the scaling law \cite{Brodsky:1973kr,Matveev:1973ra},
the cross section strongly drops with $p_{\rm lab}$ at fixed $\beta=1$, corresponding to $\Theta_{c.m.}=90\degree$ (cf. Fig.~\ref{fig:pbarn2pi-pi0}b)
which may complicate the experimental observation of the CT effects. To this end, we have also performed a calculation for $\beta=1.5$,
i.e. for $\Theta_{c.m.}=60\degree$ that is displayed in Fig.\ref{fig:sig_0deg}d. We see that the cross section is larger for $\beta=1.5$ by
an order of magnitude as compared to the case of $\beta=1$. The CT effect is still strong for $\beta=1.5$, despite the smaller momentum transfer.

\begin{figure}
  \includegraphics[scale = 0.50]{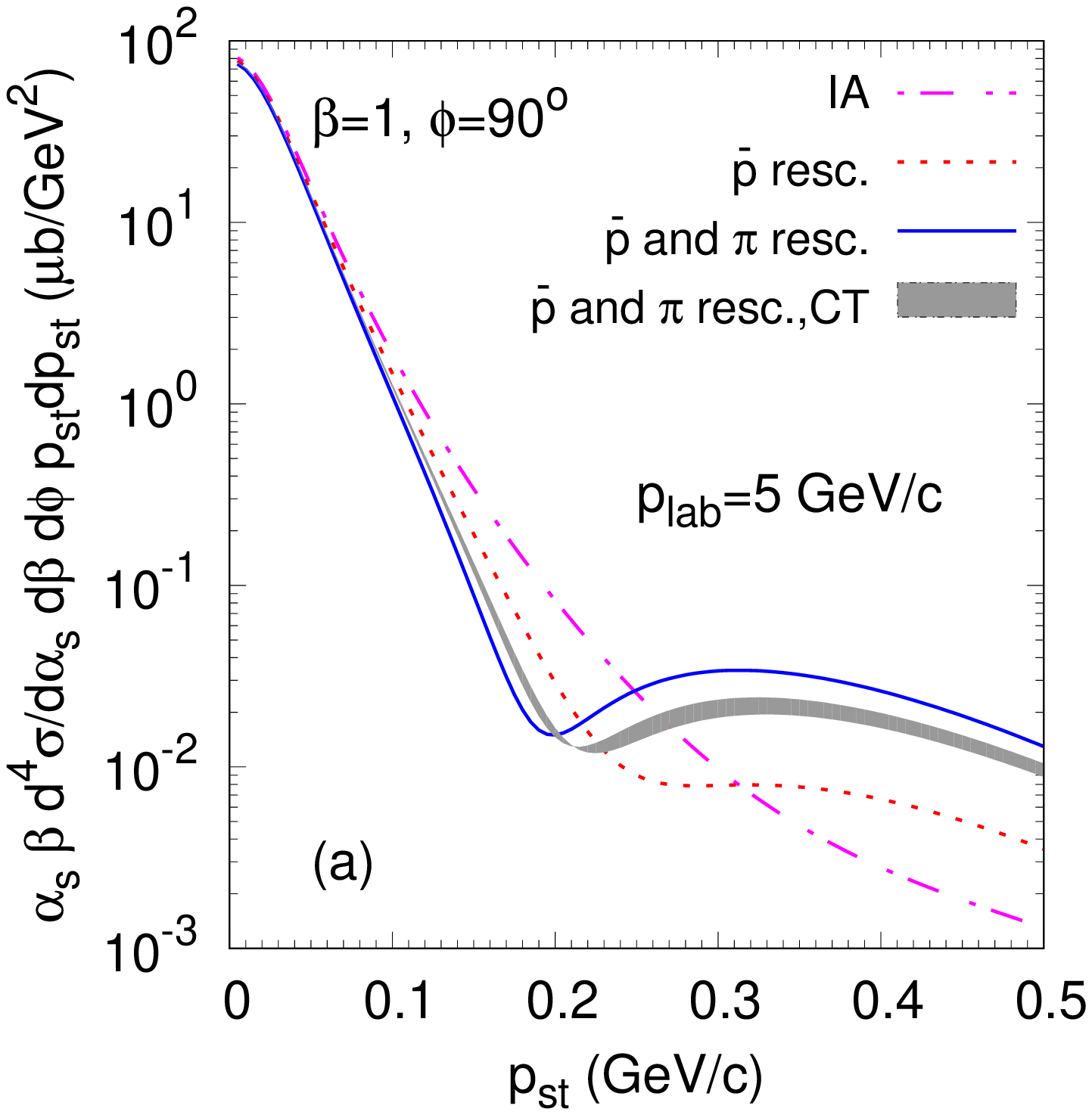}
  \includegraphics[scale = 0.50]{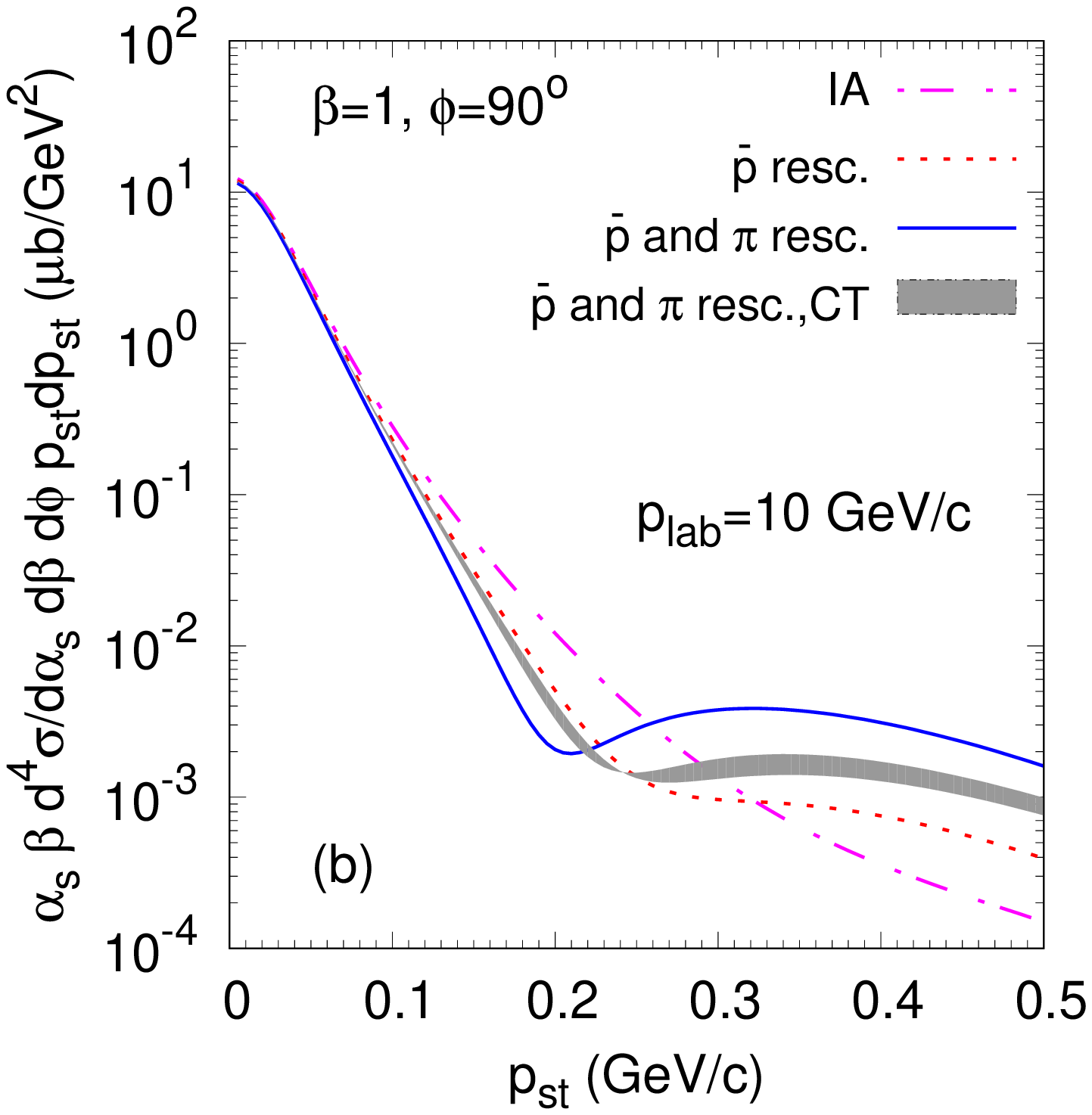}
  \includegraphics[scale = 0.50]{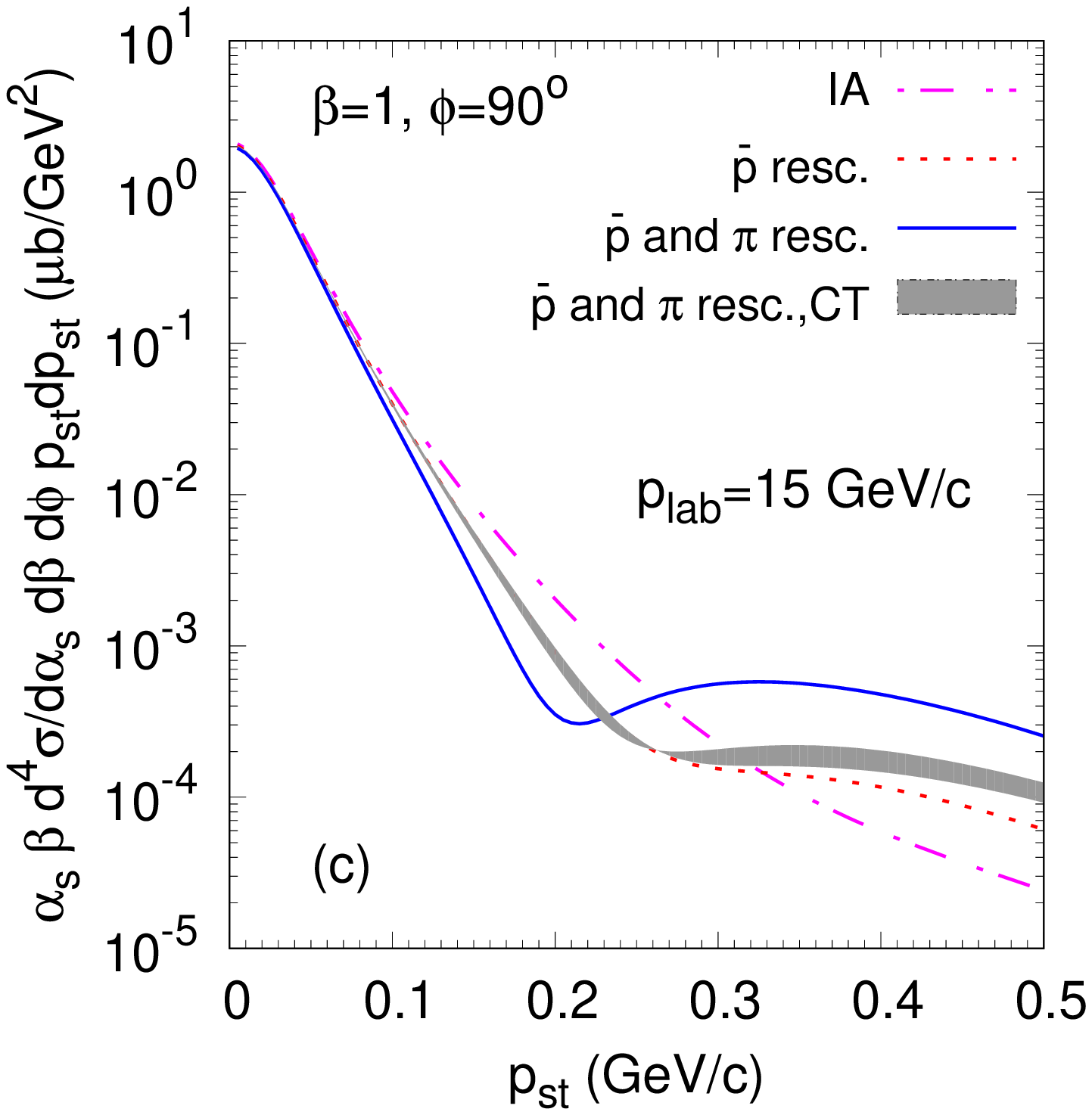}
  \includegraphics[scale = 0.50]{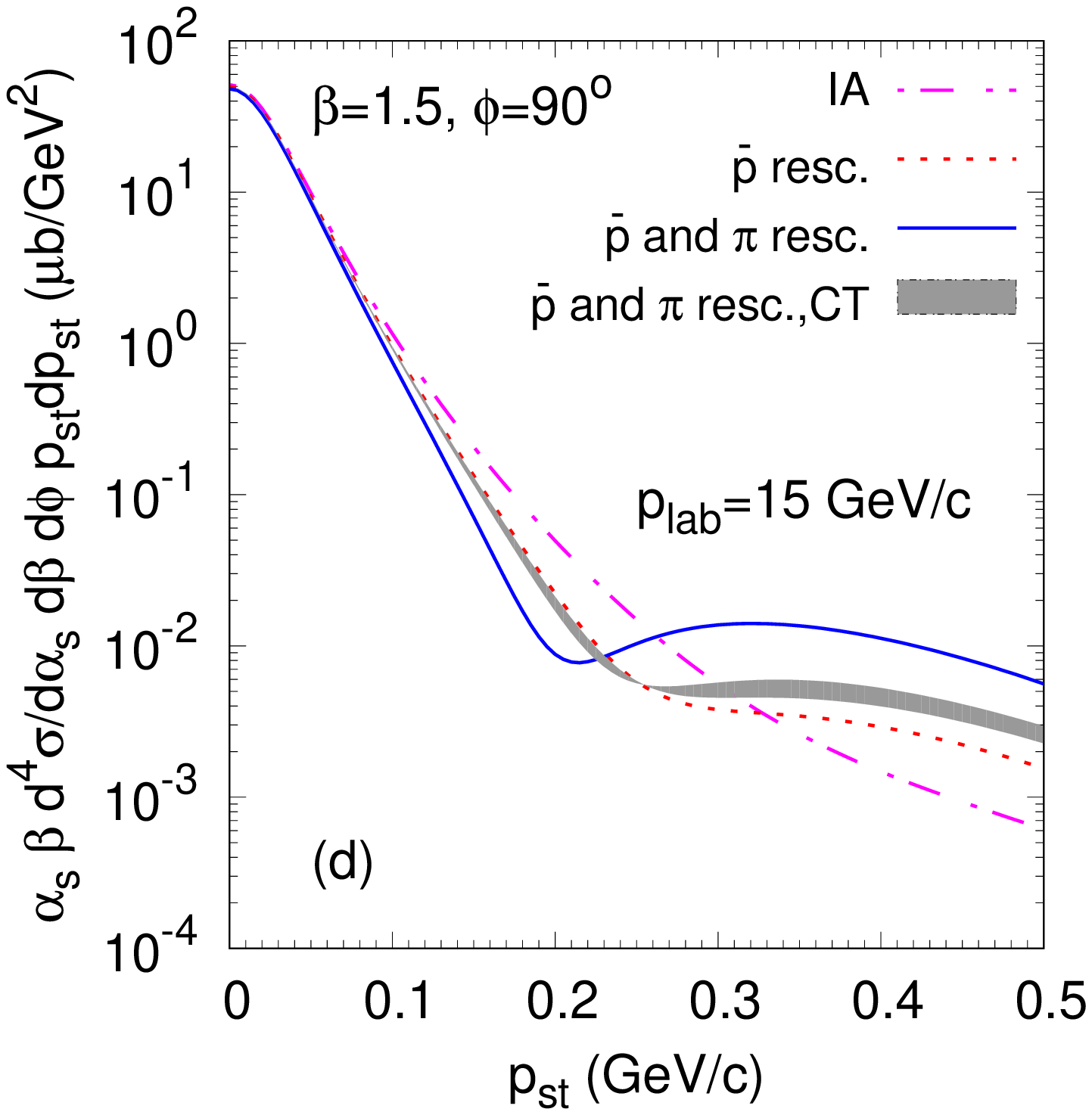}  
  \caption{\label{fig:sig_90deg} Same as Fig.~\ref{fig:sig_0deg} but for $\phi=90\degree$.}
\end{figure}
Fig.~\ref{fig:sig_90deg} shows the $p_{st}$-dependence of the four-differential cross section in the out-of-plane kinematics, $\phi=90\degree$.
Pion rescattering has much stronger effect in this case as compared to $\phi=0\degree$. This can be seen by inspecting the phases of the
exponents in Eqs.(\ref{M^(c)_coord}),(\ref{M^(d)_coord}). We see that large longitudinal (along the scattered pion) momenta of the spectator proton lead
to quickly oscillating exponents as function of longitudinal separation. Thus, relatively small longitudinal momentum transfer
$k^z \sim \Delta_{1,2}^0 \sim 0.1$ GeV/c is favored. However, the transverse momentum transfer is regulated by the momentum dependence of the elastic $\pi N$
scattering amplitude, Eq.(\ref{M_pipmp}), where the slope $B_{\pi^\pm p}$ is of the order of 7-8 GeV$^{-2}$, and so the transverse momentum
transfer $k_t \sim 0.5$ GeV/c is easily possible. For the fixed $p_{st}$ and $\beta$, the component of the spectator momentum along the scattered pion
is smaller in the out-of-plane kinematics than in the in-plane one. This leads to relatively smaller suppression of the large-$p_{st}$ cross section in the case
of the out-of-plane kinematics.

\begin{figure}
  \includegraphics[scale = 0.50]{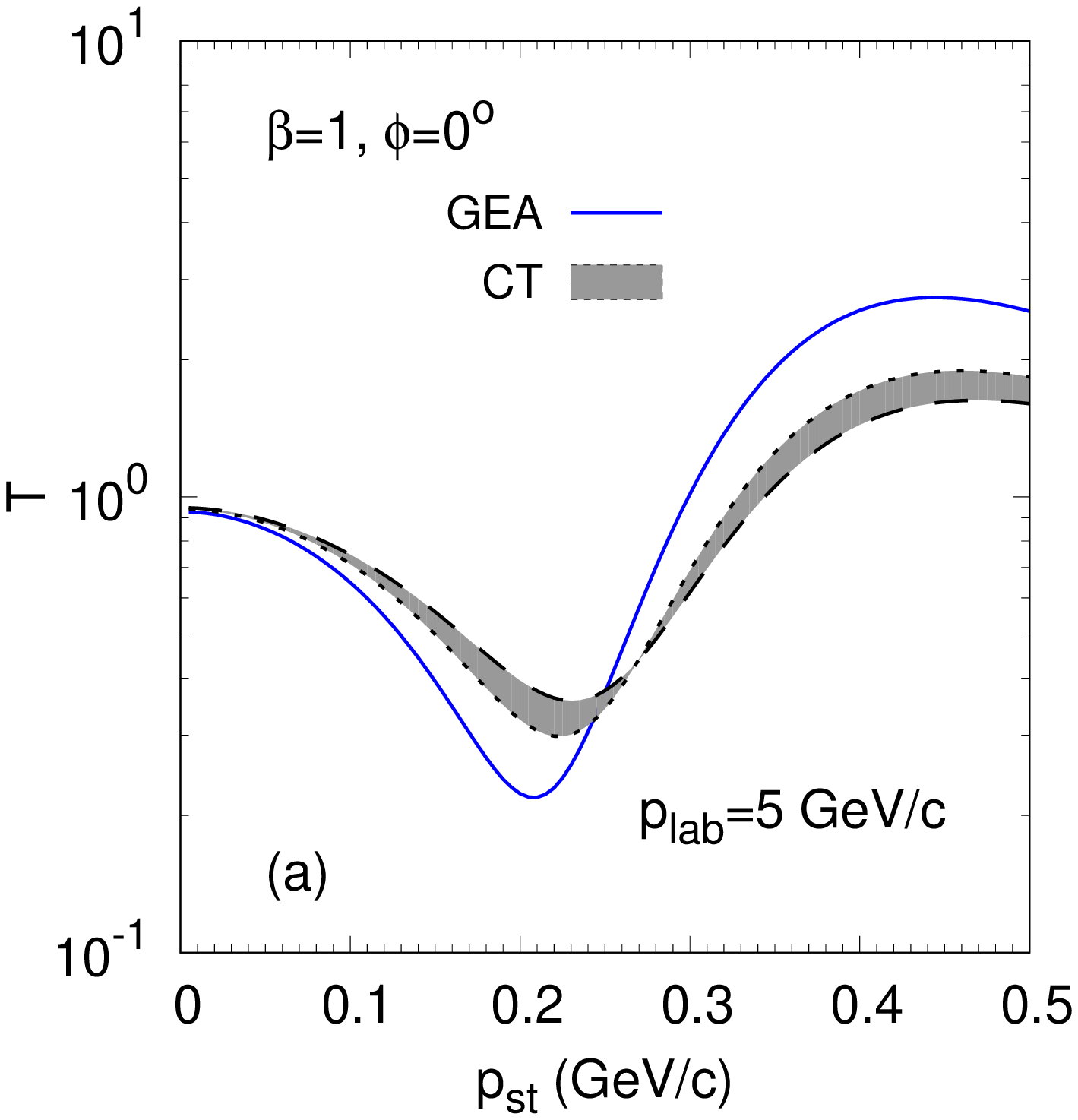}
  \includegraphics[scale = 0.50]{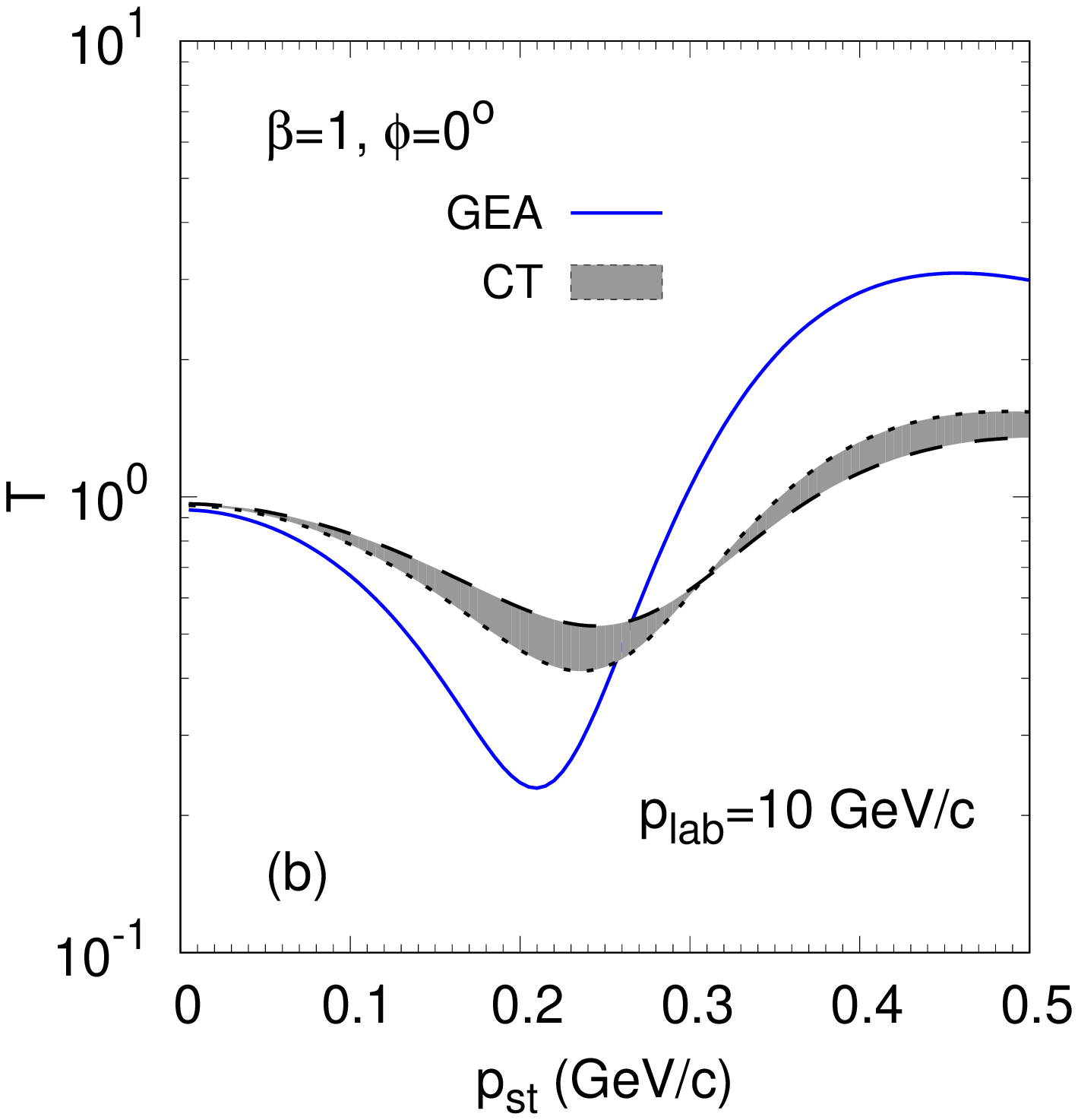}
  \includegraphics[scale = 0.50]{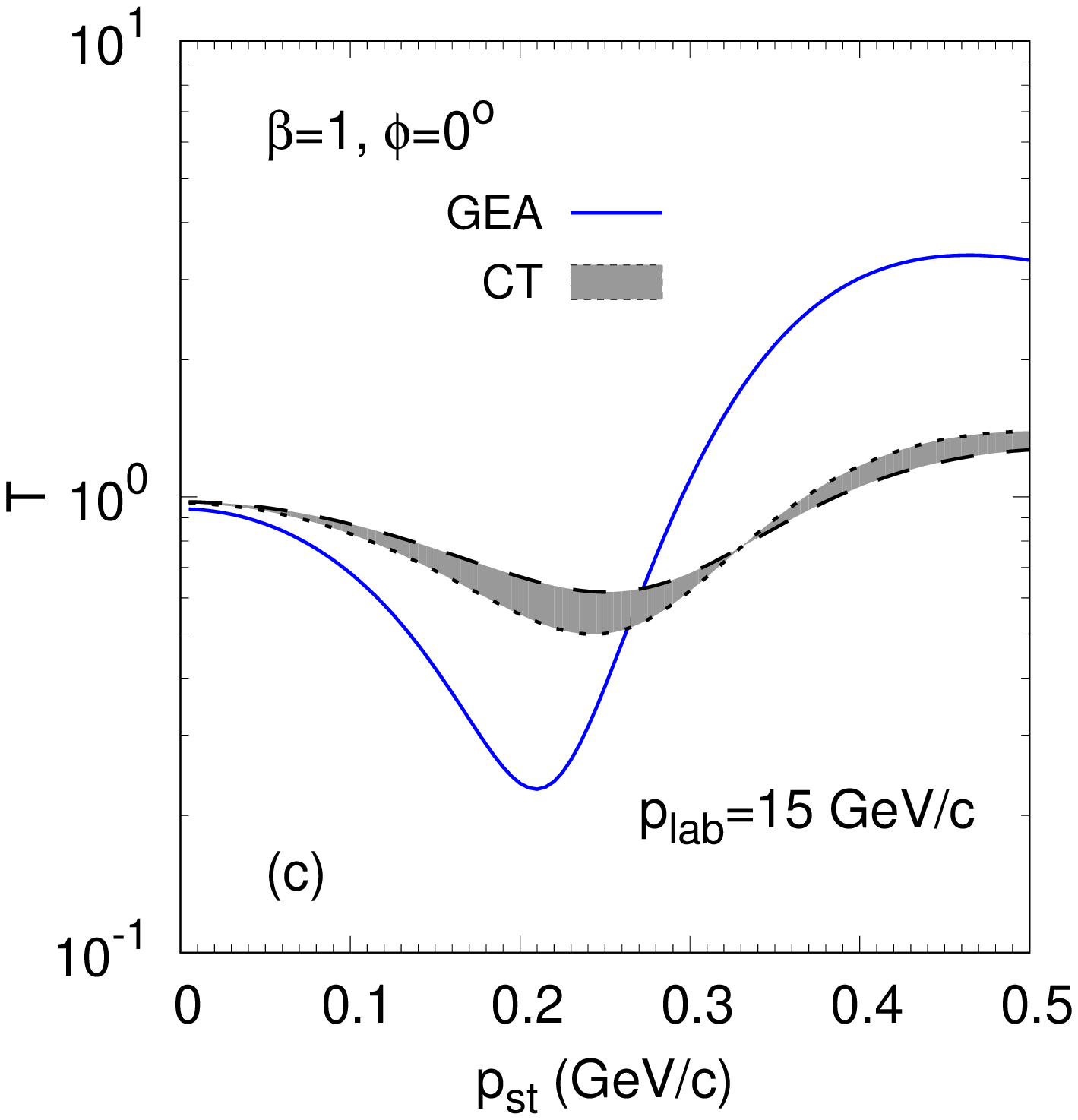}
  \includegraphics[scale = 0.50]{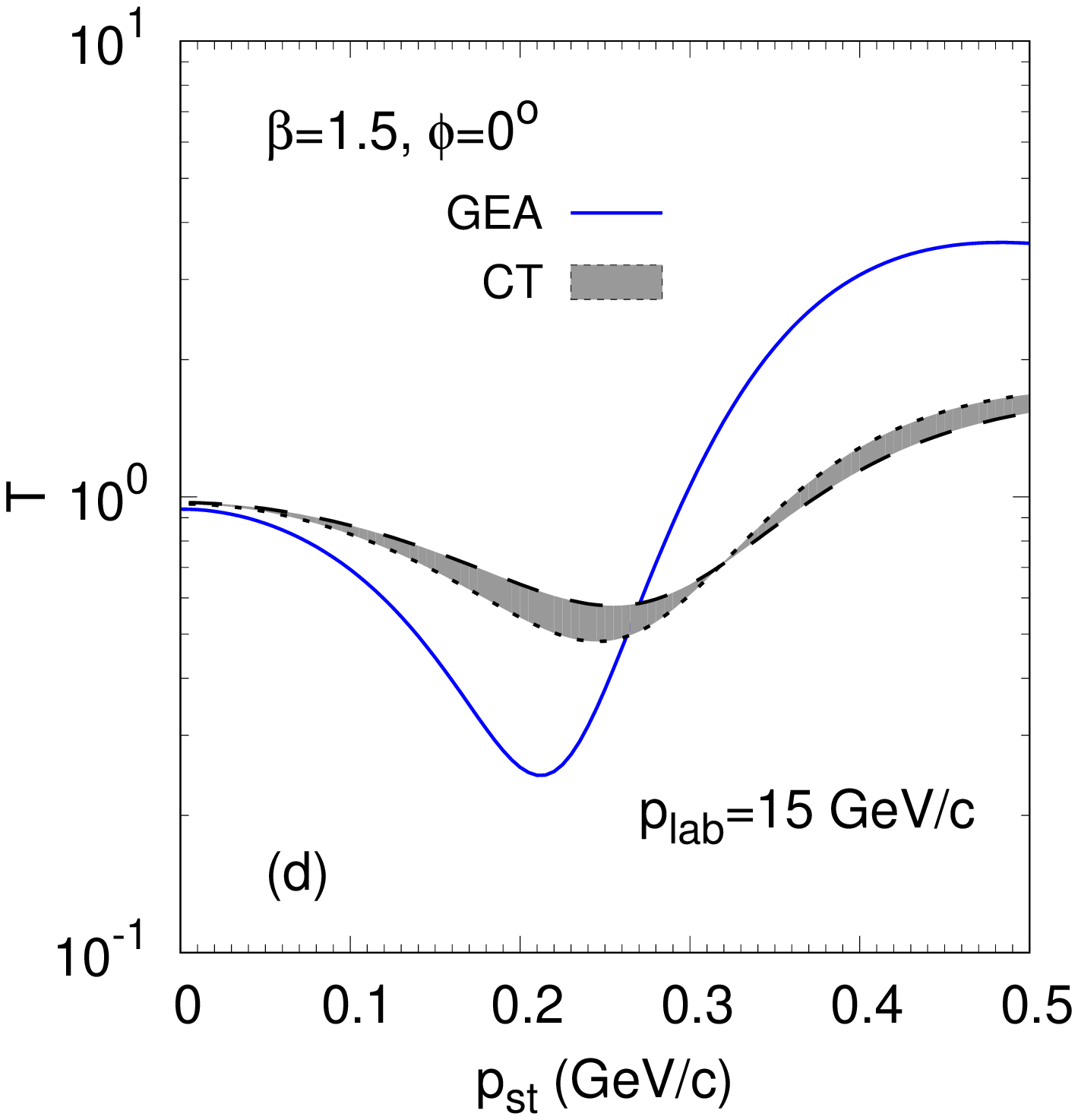}  
  \caption{\label{fig:T_0deg} Transparency ratio (see Eq.(\ref{T})) for the process $\bar p d \to \pi^- \pi^0 p$ as a function of
    transverse momentum of spectator proton for the relative azimuthal angle between $\pi^-$ and spectator proton
    $\phi=0\degree$.
    Different panels display the calculations for the different beam momenta $p_{\rm lab}$ and the LC momentum fraction
    $\beta$ carried by the $\pi^-$, as indicated. The GEA calculation is shown by the solid line.
    The calculations with CT are displayed by the grey band limited by the values of the mass denominator of the coherence length
    $\Delta M^2 = 0.7$ GeV$^2$ (dashed line) and 1.1 GeV$^2$ (dotted line). All calculations include $\bar p$ and pion rescattering amplitudes.
    The calculations are performed for $\alpha_s=1$.}
\end{figure}
\begin{figure}
  \includegraphics[scale = 0.50]{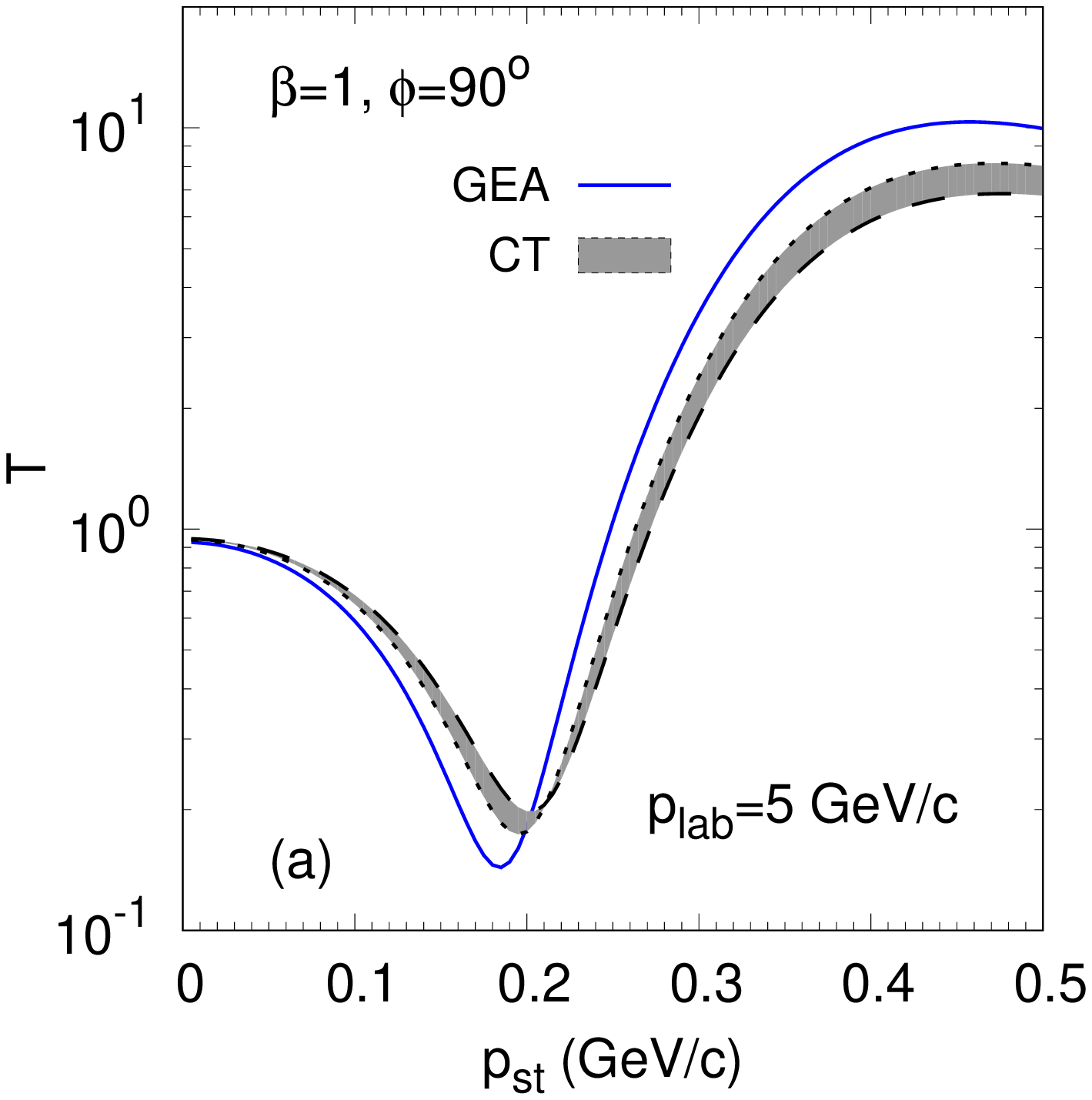}
  \includegraphics[scale = 0.50]{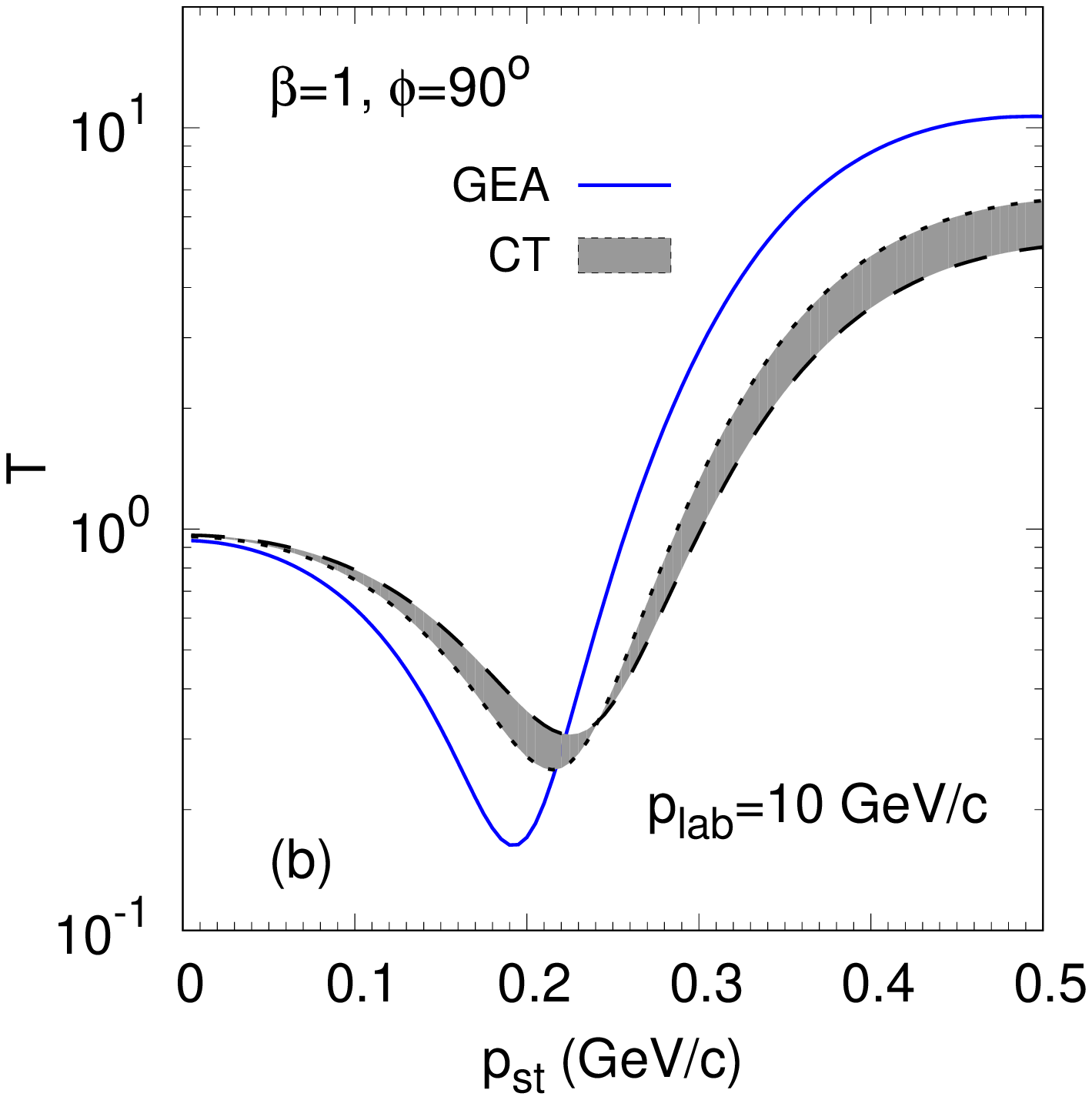}
  \includegraphics[scale = 0.50]{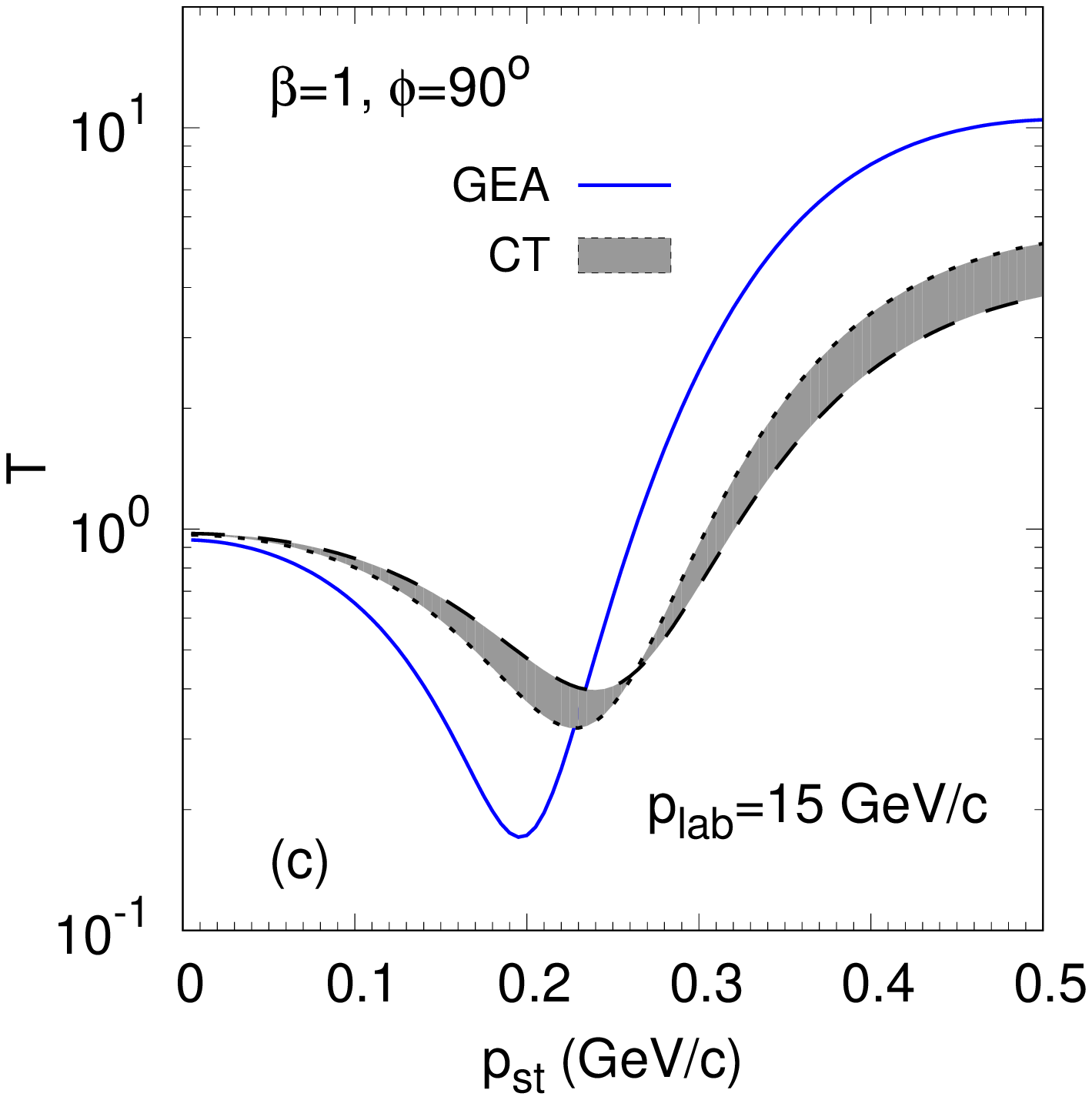}
  \includegraphics[scale = 0.50]{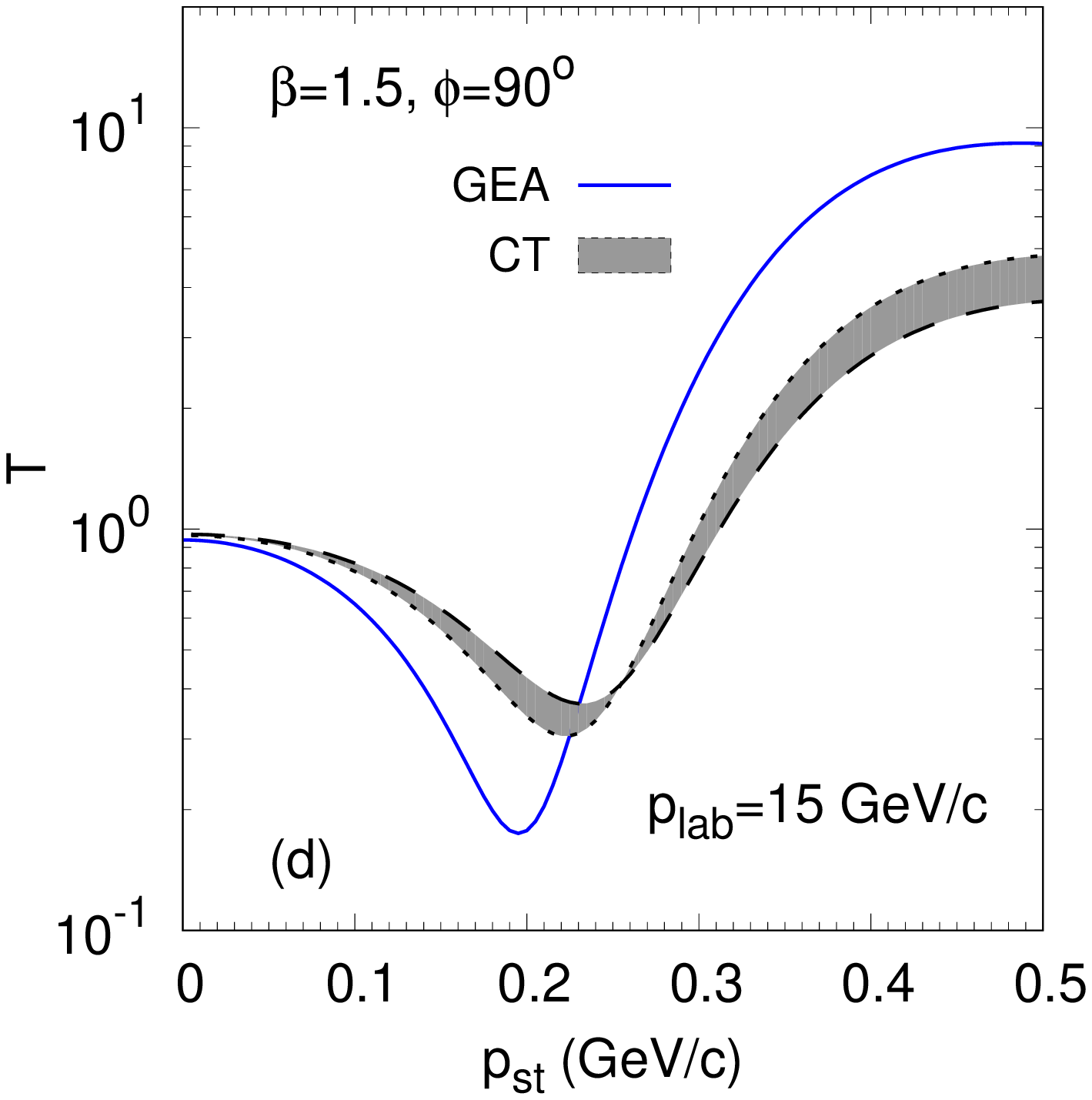}  
  \caption{\label{fig:T_90deg} Same as Fig.~\ref{fig:T_0deg} but for $\phi=90\degree$.}
\end{figure}
Figs.~\ref{fig:T_0deg},\ref{fig:T_90deg} show the transparency ratio $T$ vs spectator transverse momentum $p_{st}$ for the in-plane and out-of-plane
kinematics, respectively. At $p_{st} \ltsim 0.3$ GeV/c we observe nuclear absorption, i.e. $T < 1$. The absorptive behaviour arises due to the
interference term between the IA and rescattering amplitudes, similar to the previous GEA studies \cite{Frankfurt:1996uz}. The interference term
governs the decrease of $T$ with growing transverse momentum of the spectator at $p_{st} \ltsim 0.2$ GeV/c. At larger $p_{st}$'s the squared rescattering
amplitudes become dominant leading to the increase of the transparency ratio, especially strong for the out-of-plane kinematics.
CT reduces the effect of rescattering amplitudes both at low and high transverse momenta and thus smooths down the structures
in the $p_{st}$-dependence of the transparency ratio.
Since the absolute values of the rescattering amplitudes calculated
with CT are generally smaller than those in GEA, the transition from the absorption to the rescattering
regime takes place at larger transverse momenta of the spectator. 
The effect of CT is more pronounced for the in-plane kinematics that can be again explained by quickly
oscillating exponents in Eqs.(\ref{M^(c)_coord}),(\ref{M^(d)_coord}) as functions of longitudinal separation.
In this case the integrals are dominated by small longitudinal separations where 
CT is stronger.

\begin{figure}
  \begin{tabular}{ll}
    \includegraphics[scale = 0.50]{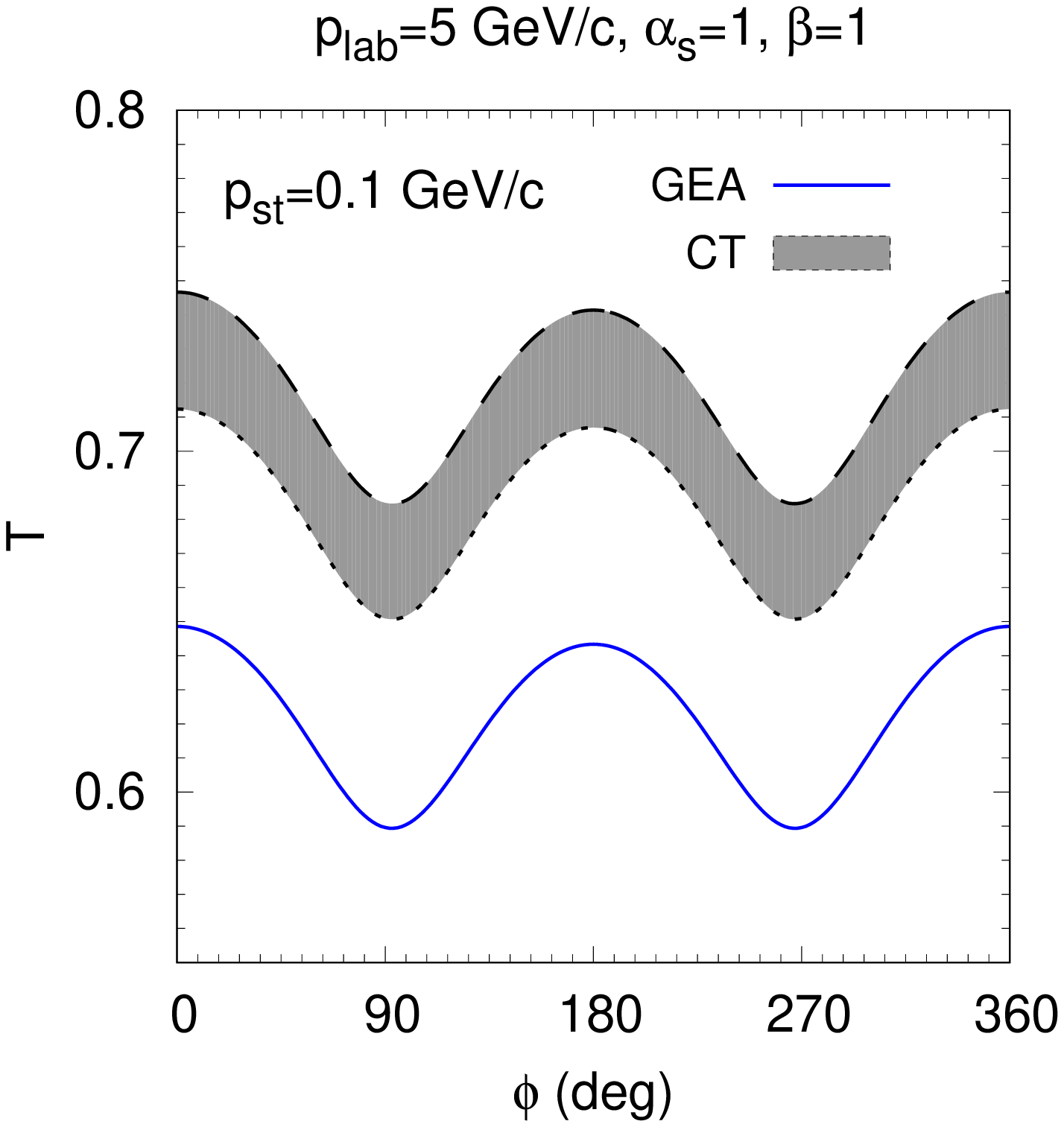} &
    \includegraphics[scale = 0.50]{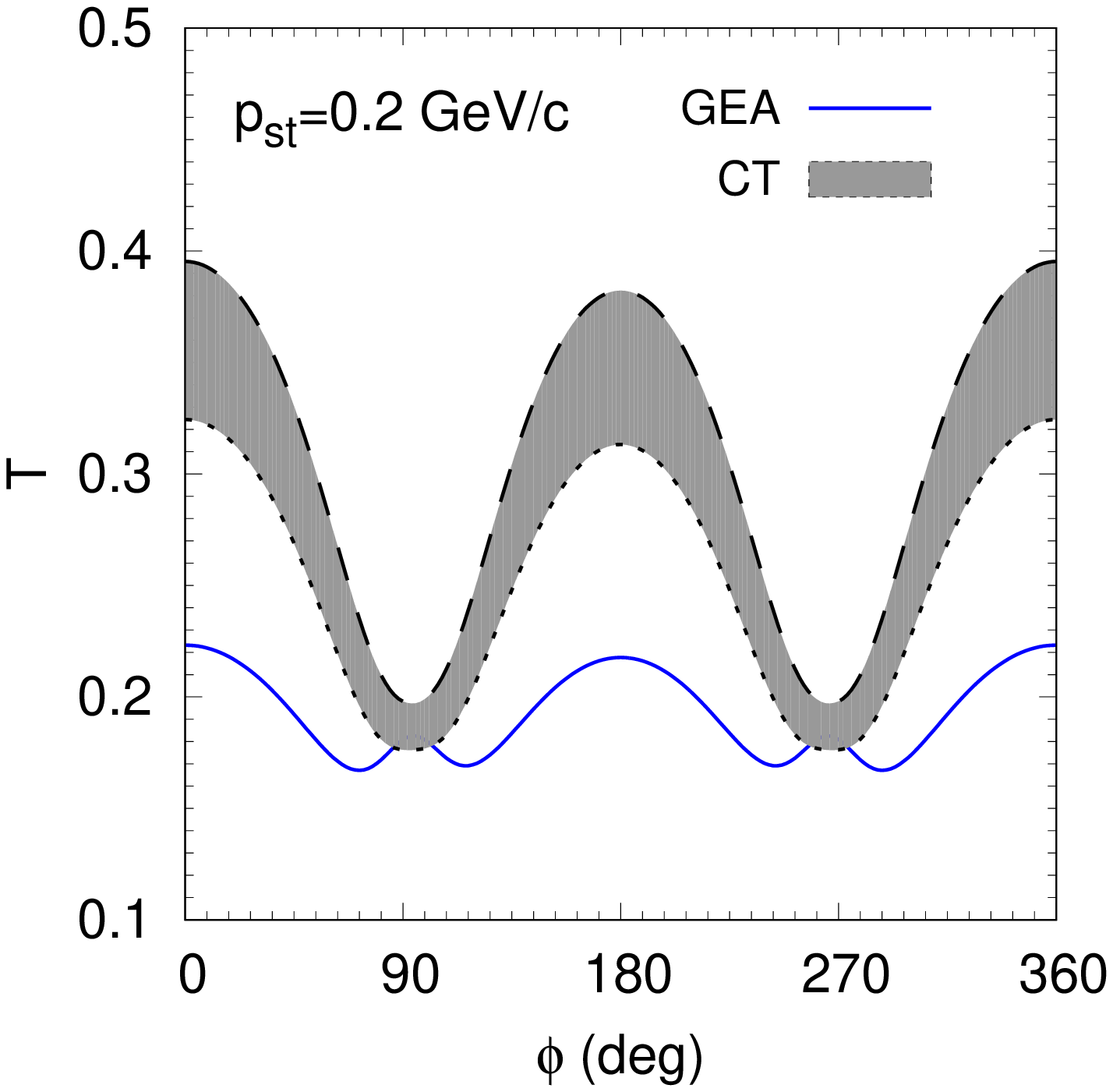}\\
    \includegraphics[scale = 0.50]{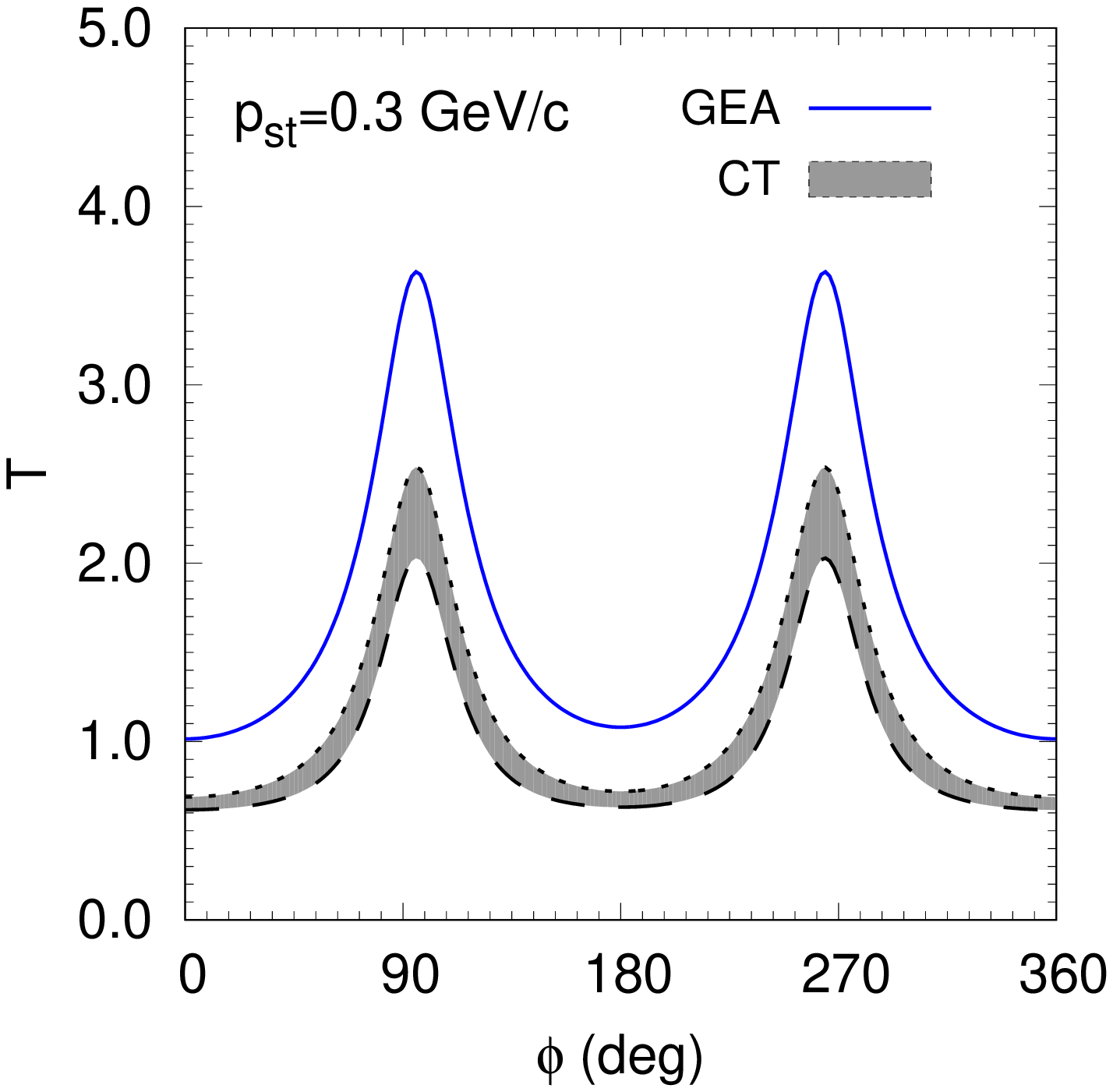} &
    \includegraphics[scale = 0.50]{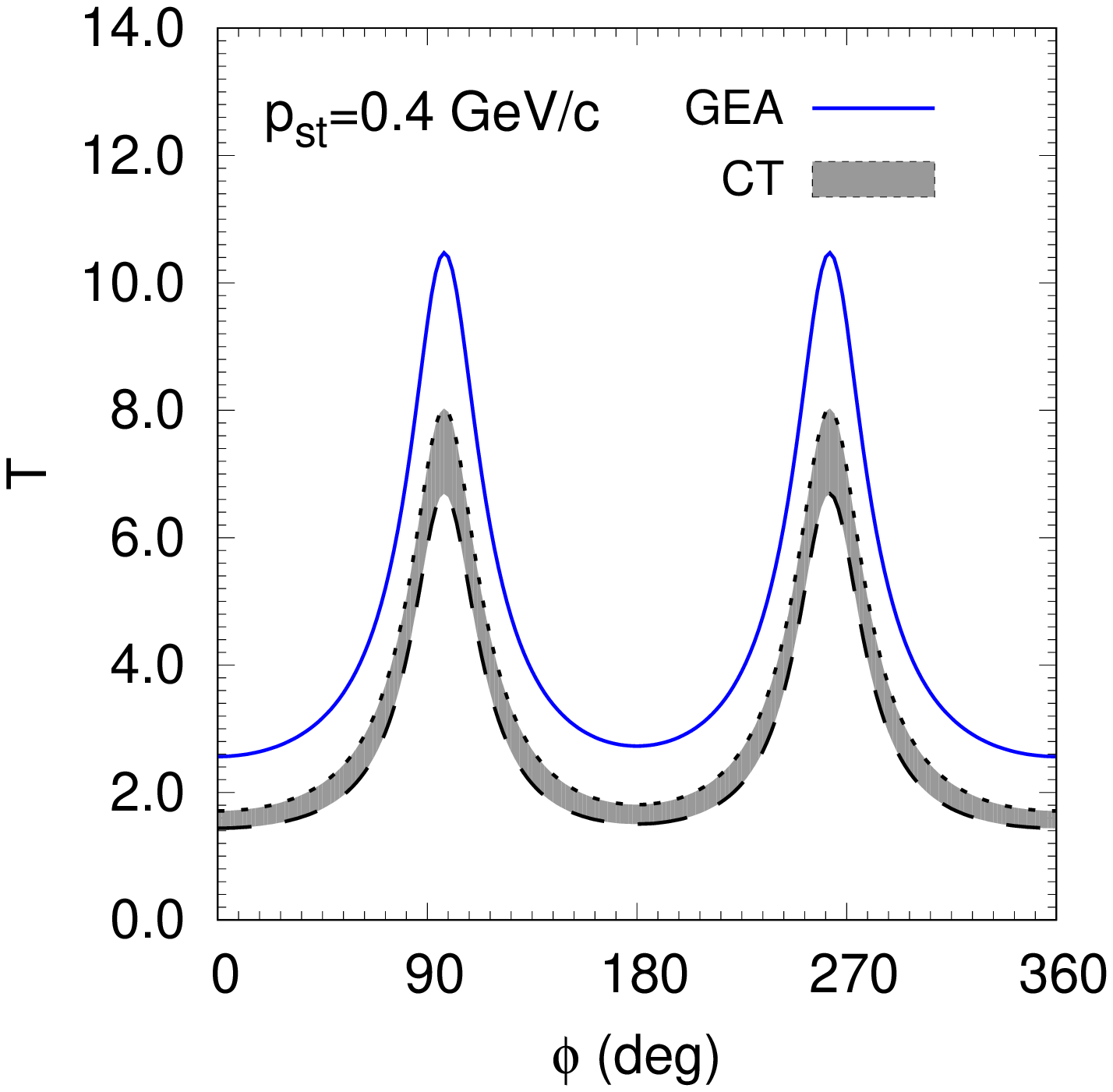}\\
  \end{tabular} 
  \caption{\label{fig:T_phiDep_5gevc} Transparency ratio (see Eq.(\ref{T})) for the process $\bar p d \to \pi^- \pi^0 p$
    at the beam momentum of 5 GeV/c, $\alpha_s=1$, $\beta=1$ as a function of
    the relative azimuthal angle between $\pi^-$ and spectator proton for the different transverse momenta $p_{st}$ of the spectator proton.
    The GEA calculation is shown by the solid line.
    The calculations with CT are displayed by the grey band limited by the values of the mass denominator of the coherence length
    $\Delta M^2 = 0.7$ GeV$^2$ (dashed line) and 1.1 GeV$^2$ (dotted line). All calculations include $\bar p$ and pion rescattering amplitudes.}
\end{figure}
\begin{figure}
  \begin{tabular}{ll}
    \includegraphics[scale = 0.50]{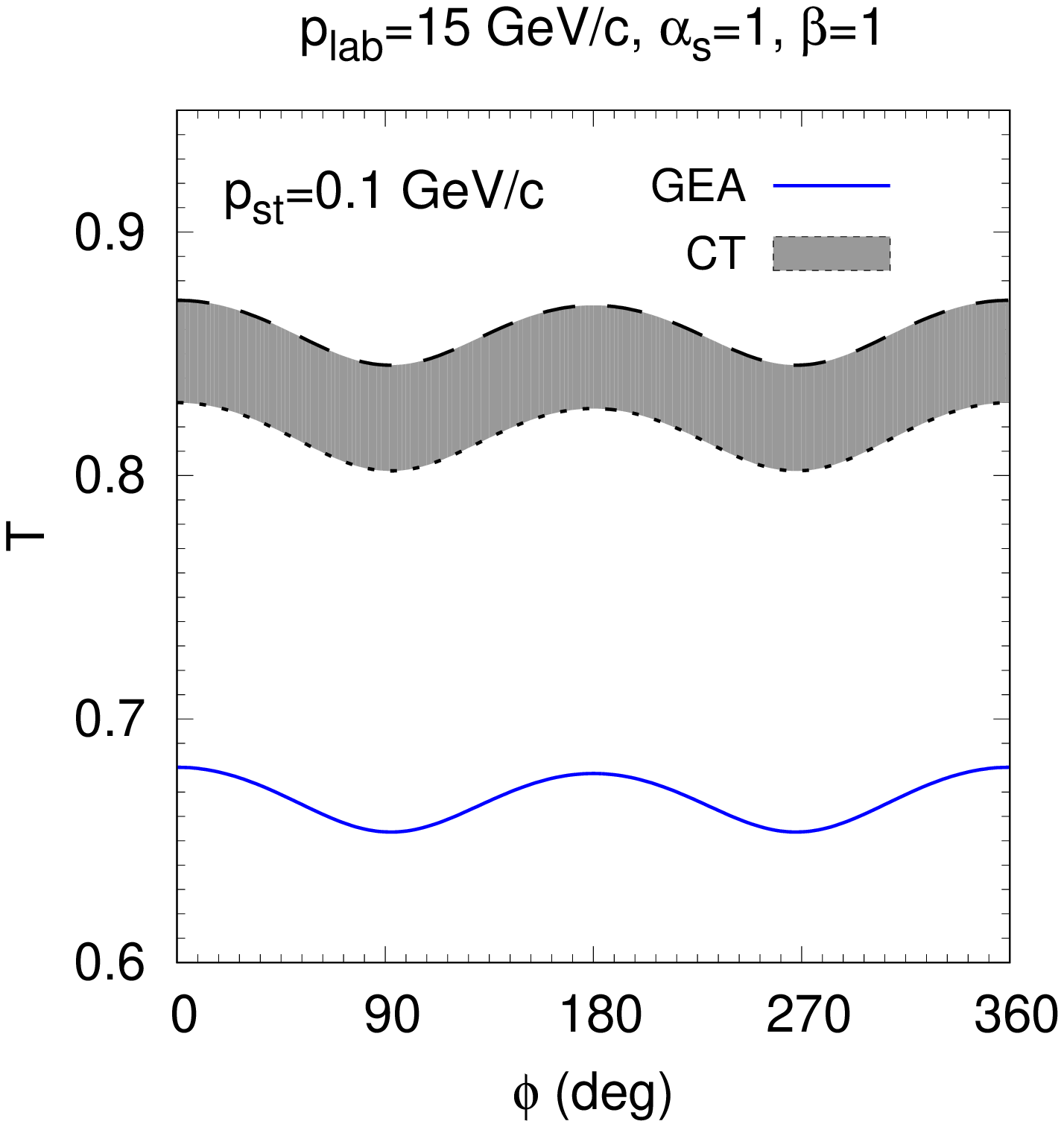} &
    \includegraphics[scale = 0.50]{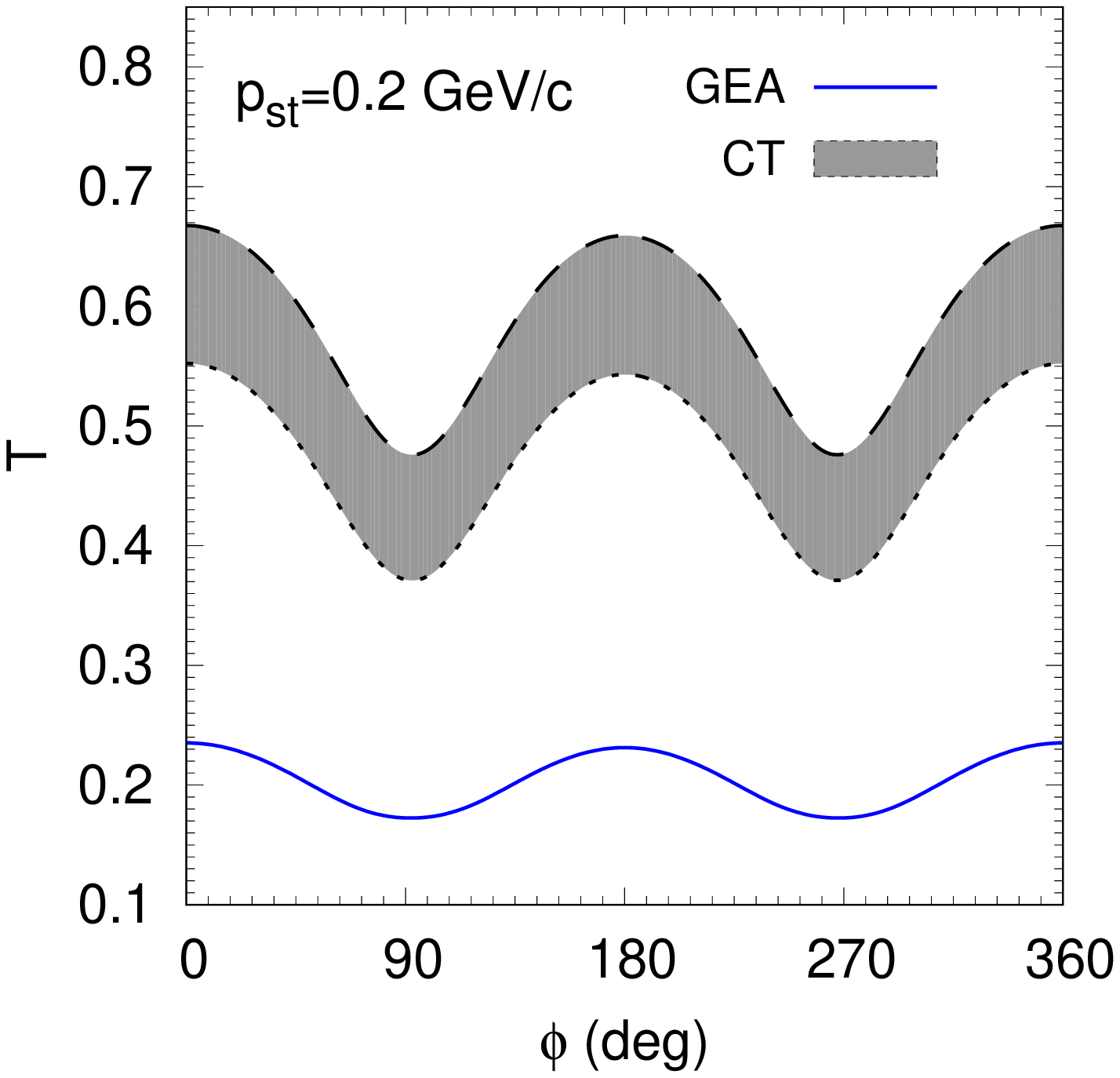}\\
    \includegraphics[scale = 0.50]{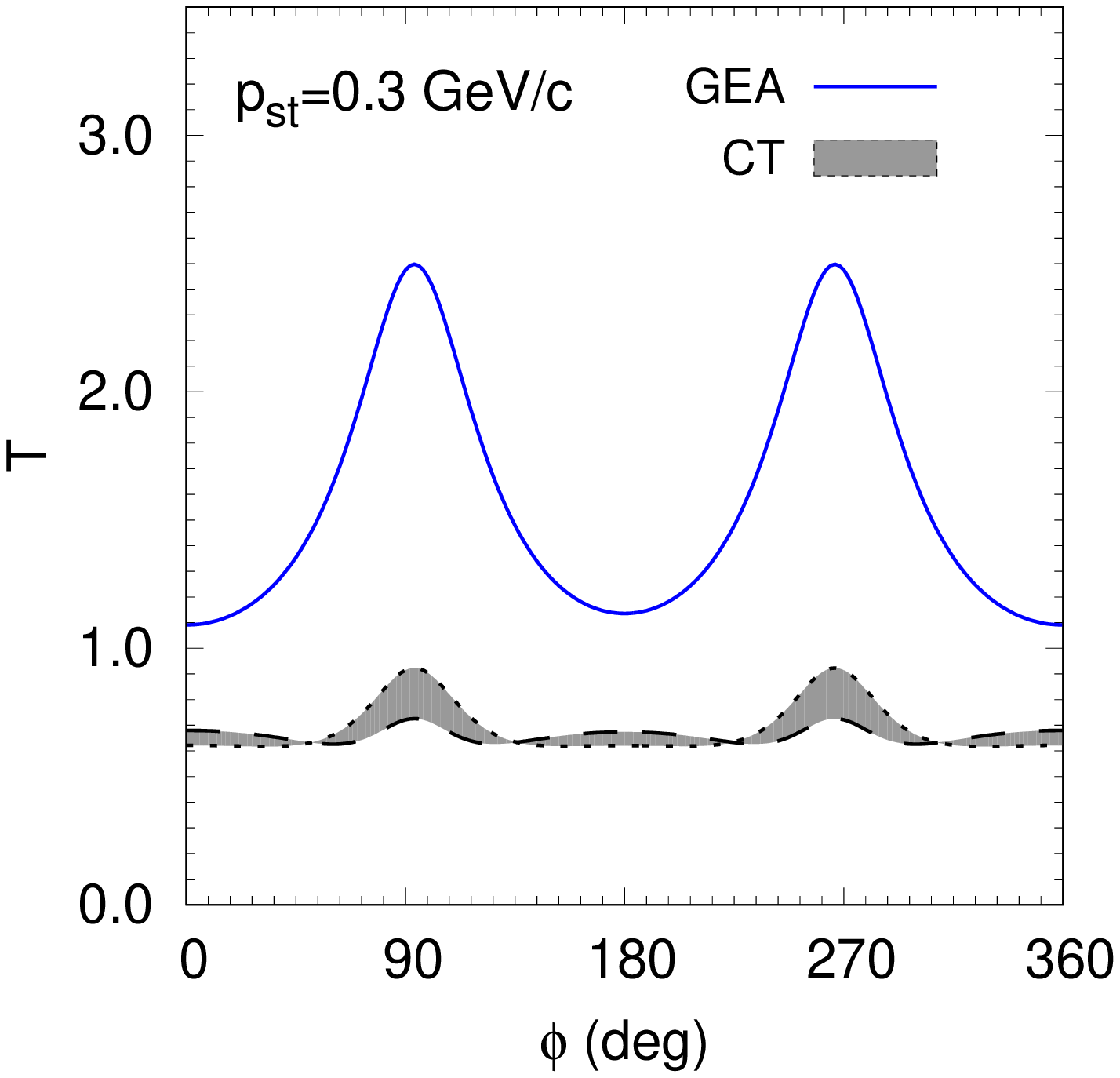} &
    \includegraphics[scale = 0.50]{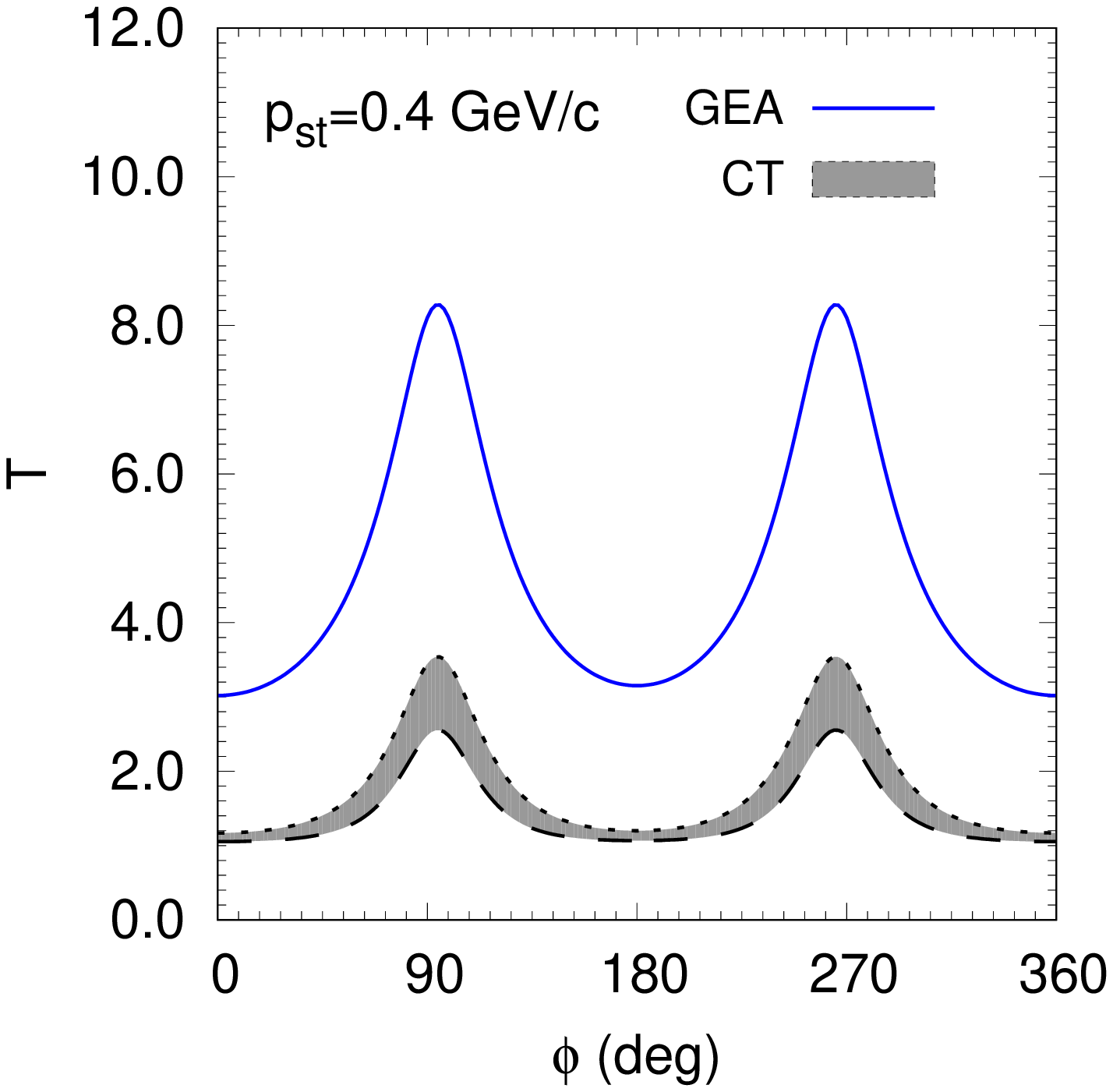}\\
  \end{tabular} 
  \caption{\label{fig:T_phiDep_15gevc} Same as Fig.~\ref{fig:T_phiDep_5gevc} but for $p_{\rm lab}=15$ GeV/c.}
\end{figure}
Figs.~\ref{fig:T_phiDep_5gevc} and \ref{fig:T_phiDep_15gevc} show the azimuthal angle dependence of the transparency ratio $T$ for
for low and high beam momenta, respectively. For $p_{st} \ltsim 0.2$ GeV/c, we observe the in-plane ($\phi=0\degree$ and $180\degree$)
enhancement, while at higher transverse momenta -- the out-of-plane ($\phi=90\degree$ and $270\degree$) enhancement sets in.
This is due to a larger pion rescattering amplitude in the case when the momenta of the scattered pion and spectator proton are orthogonal.
At small transverse momentum of the spectator proton the destructive interference of the pion rescattering amplitude
and the IA amplitude leads to a larger absorption for $\phi=90\degree$ and $270\degree$
while at large transverse momentum the dominating pion rescattering amplitude squared
leads to the enhanced production for the same relative azimuthal angles. At the intermediate transverse momenta,  
a more complicated shape of the azimuthal dependence emerges in the GEA calculations at smaller beam momenta
(see Fig.~\ref{fig:T_phiDep_5gevc} for $p_{st} = 0.2$ GeV/c). The CT effect on the azimuthal dependence of $T$ is more pronounced
at higher beam momenta. We note that the choice of the LC variable $\beta$ (1 or 1.5) has practically no influence on the
$\phi$-dependence of the transparency ratio.
At smaller (larger) transverse momenta of the spectator, 
CT leads to stronger (weaker) variation of the transparency ratio
with the azimuthal angle as compared to the GEA calculation.  
Similar behaviour was found in ref. \cite{Frankfurt:1996uz}
for the case of $d(p,2p)n$ process (see Fig.~7 of ref. \cite{Frankfurt:1996uz}).

\section{Monte-Carlo simulations}
\label{MC}

We have performed Monte-Carlo (MC) simulations primarily aimed to evaluate event rates at PANDA.
Two types of MC simulations have been performed.

In the first simulation run, we have calculated the integrated cross section
applying the kinematic cuts on the produced particles. The integrated cross section has been calculated as follows:
\begin{equation}
  \sigma_{\bar p d \to \pi_1^- \pi_2^0 p} =  
  \frac{(2\pi)^4}{4p_{\rm lab}m_d} \Phi_3  \langle \overline{|M|^2} {\cal F}(\bvec{k}_1,\bvec{k}_2,\bvec{p}_s) \rangle~,          \label{sigma}
\end{equation}
where 
\begin{equation}
  \Phi_3 = \int d\Phi_3     \label{Phi_3}
\end{equation}
is the integrated phase space volume (see Eq.(\ref{dPhi_3})), ${\cal F}$ is the function defined so that
${\cal F}=1(0)$ if the outgoing momenta $\bvec{k}_1,\bvec{k}_2,\bvec{p}_s$ are within (out of) the cutting region.
The integrated phase space volume $\Phi_3$ has been calculated numerically using Dalitz-type expressions
(cf. \cite{Montanet:1994xu}). The angular brackets $\langle \ldots \rangle$ in  Eq.(\ref{sigma}) denote the averaging
over phase space volume which has been calculated by MC sampling momenta $\bvec{k}_1,\bvec{k}_2,\bvec{p}_s$ with
probability $dP \propto d\Phi_3$.
In that way we obtained the integrated cross section collected in Table \ref{tab:Sigma_int}.
\begin{table}[htb]
  \caption{\label{tab:Sigma_int} Integrated cross sections of the process $\bar p d \to \pi^- \pi^0 p_s$  
  in the region  $p_{st}=0-0.5$ GeV/c, $\alpha_s=0.9-1.1$
  and the numbers of events per day for the high-luminosity mode of PANDA, $L=2 \cdot 10^{32}~\mbox{cm}^{-2} \mbox{s}^{-1}$.
  Listed are the values calculated with CT, $\Delta M^2=0.7$ GeV$^2$ and (in parentheses) without CT. }
  \begin{center}
    \begin{tabular}{|c|c|c|c|}
    \hline
    $p_{\rm lab}$, GeV/c    &   $\beta$   &  $\sigma_{\rm int}$,  pb   &   $\mbox{events}/\mbox{day}$    \\
    \hline
    5                                        &  0.9-1.1    & 8039 (7558)                         &    138920 (130597)      \\
   10                                       &  0.9-1.1    & 1309 (1179)                         &      22616 (20365)        \\
   15                                       &   0.9-1.1   & 229  (202)                             &        3965 (3496)      \\
   15                                       &   1.4-1.6   & 3429 (3052)                          &      59251 (52744)  \\
\hline
    \end{tabular}
  \end{center}
\end{table}
We see that for $\beta=1\pm0.1$ the integrated cross sections are of the order of nb which allows to collect 
statistics of the order of $10^5$ and $10^3$ events per day at 5 and 15 GeV/c, respectively.
Choosing the range $\beta=1.5\pm0.1$ leads to an order of magnitude larger cross sections at 15 GeV/c.

In the second simulation run, we performed the sampling of events with probability
$dP \propto  {\cal F}(\bvec{k}_1,\bvec{k}_2,\bvec{p}_s) \overline{|M|^2} d\Phi_3$.    
At $p_{\rm lab}=15$ GeV/c, we simulated 40000 events for $\beta=0.9-1.1$ and 600000 events for $\beta=1.4-1.6$ with the full matrix element
(including all rescattering amplitudes).
As one sees from Table~\ref{tab:Sigma_int}, this statistics corresponds to $\sim 10$ days of measurements.
Since the $p_{st}$-integrated cross sections
are practically insensitive to the rescattering amplitudes, the transparency ratio can be thus calculated as the ratio of the differential probabilities
obtained with full and IA matrix elements:
\begin{equation}
  T=\frac{dP^{\rm full}/dp_{st}}{dP^{\rm IA}/dp_{st}}~.    \label{T_MC}
\end{equation}
The distribution $dP^{\rm full}/dp_{st}$ is supposed to be measurable in the experiment, while $dP^{\rm IA}/dp_{st}$
is a pure theoretical distribution.
Thus, in the latter case the number of events is, in principle, limited by the available computing resources only.
In order to calculate the transparency ratio (\ref{T_MC}), we simulated 200000 and 600000 events with the IA matrix element for $\beta=0.9-1.1$
and $\beta=1.4-1.6$, respectively. This statistics is enough for the present exploratory studies.

\begin{figure}
  \includegraphics[scale = 0.50]{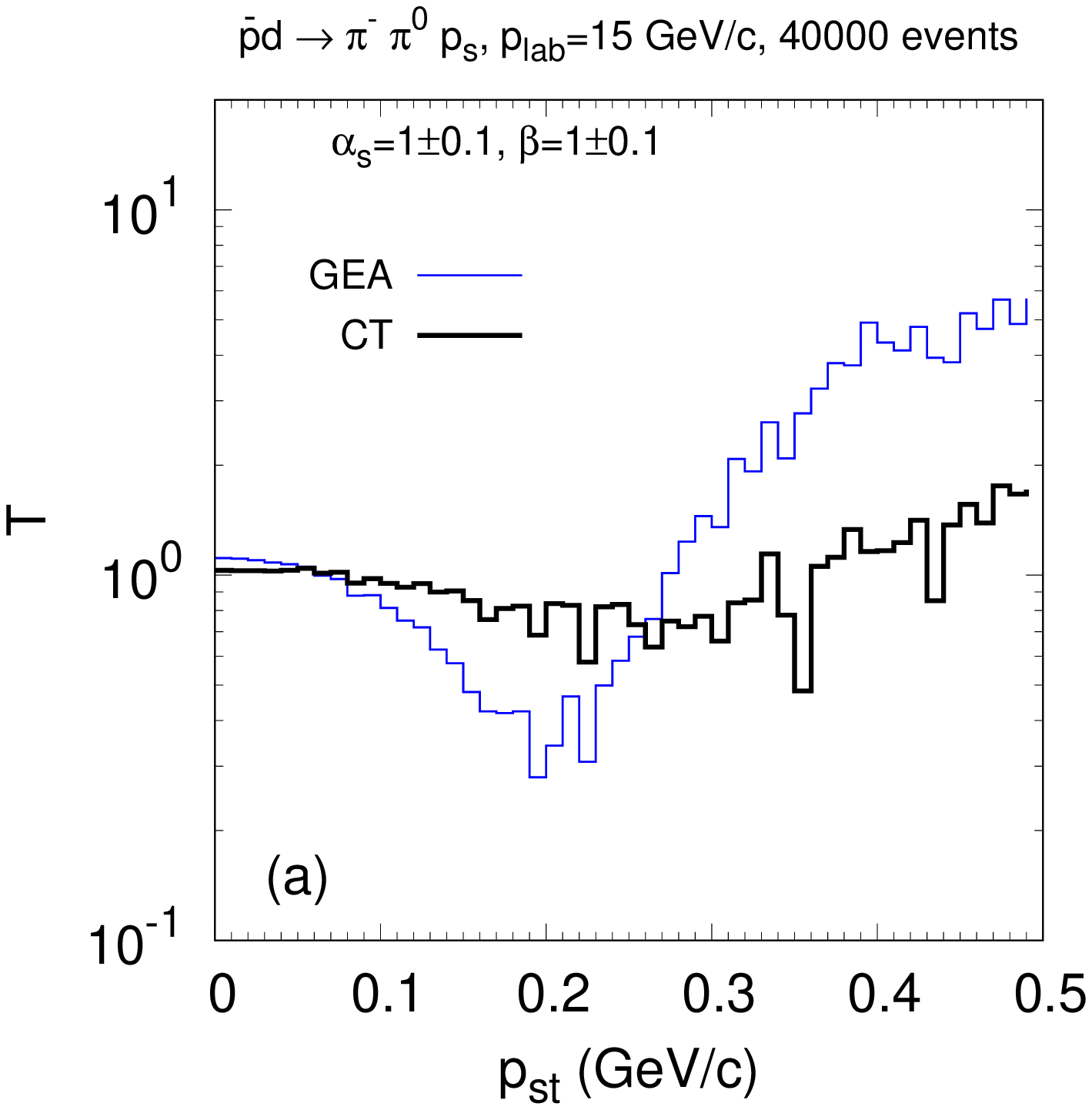}
  \includegraphics[scale = 0.50]{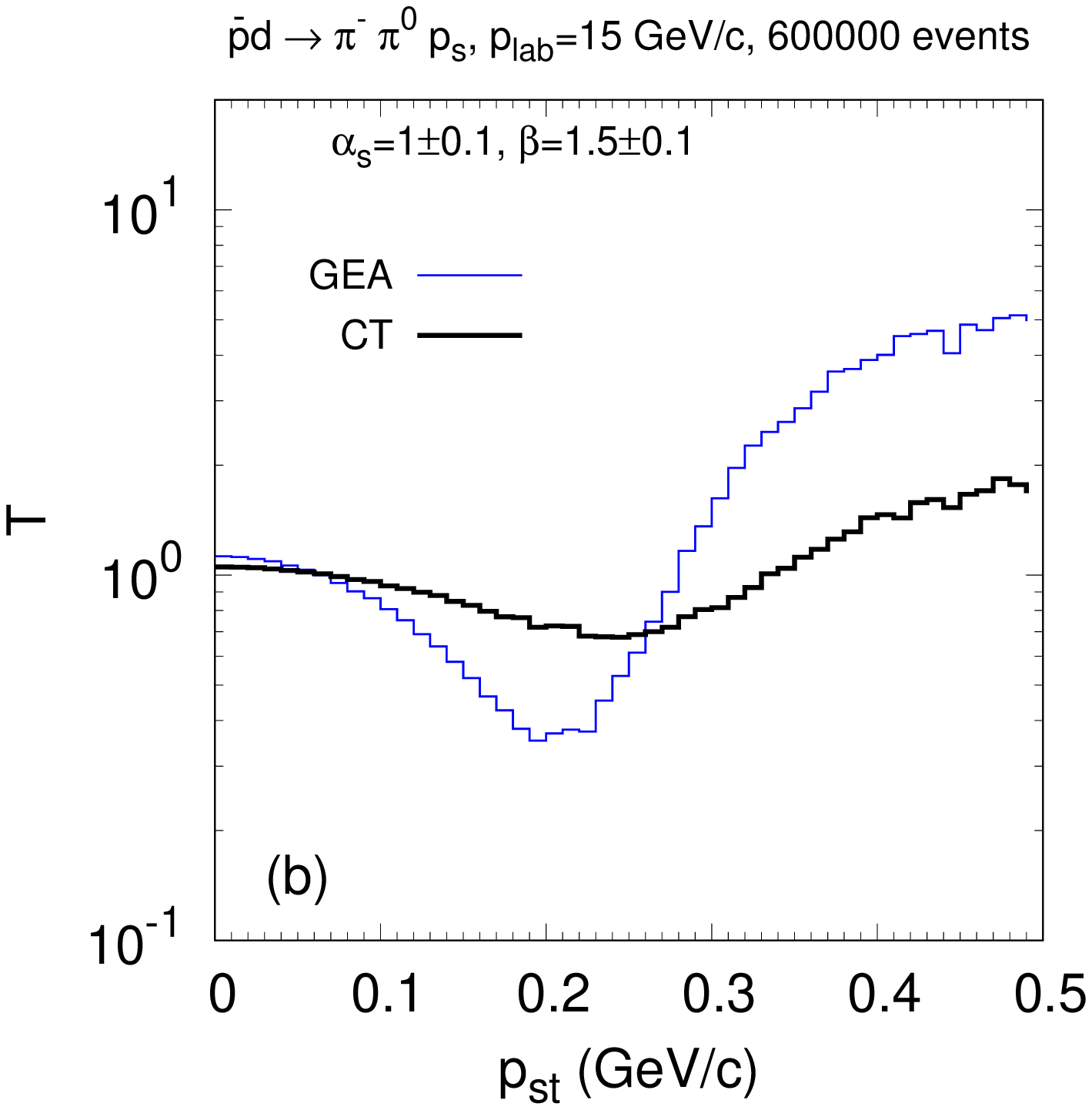}
  \caption{\label{fig:Evdist_15gevc} Transparency ratio calculated in the MC simulation (see Eq.(\ref{T_MC})) for the process $\bar p d \to \pi^- \pi^0 p$
    at the beam momentum of 15 GeV/c, $\alpha_s=1\pm0.1$ 
    as a function of the transverse momentum $p_{st}$ of the spectator proton.
    Panels (a) and (b) show, respectively the kinematics with $\beta=1\pm0.1$ and $\beta=1.5\pm0.1$.
    The GEA calculation is shown by the thin blue line.
    The calculation with CT for $\Delta M^2 = 0.7$ GeV$^2$ is shown by the thick black line.}
\end{figure}
Fig.~\ref{fig:Evdist_15gevc} shows the spectator transverse momentum dependence of the transparency ratio (\ref{T_MC}) at $p_{\rm lab}=15$ GeV/c.
We see that the MC results agree with the results of direct calculations, Figs.~\ref{fig:T_0deg}c,d and \ref{fig:T_90deg}c,d
(some difference between the simulated and the direct results is mostly due to averaging over relative azimuthal angle in the MC
simulations). It is also evident that the difference between the GEA and the CT calculations becomes observable
even with our modest statistics. 
\begin{figure}
  \includegraphics[scale = 0.50]{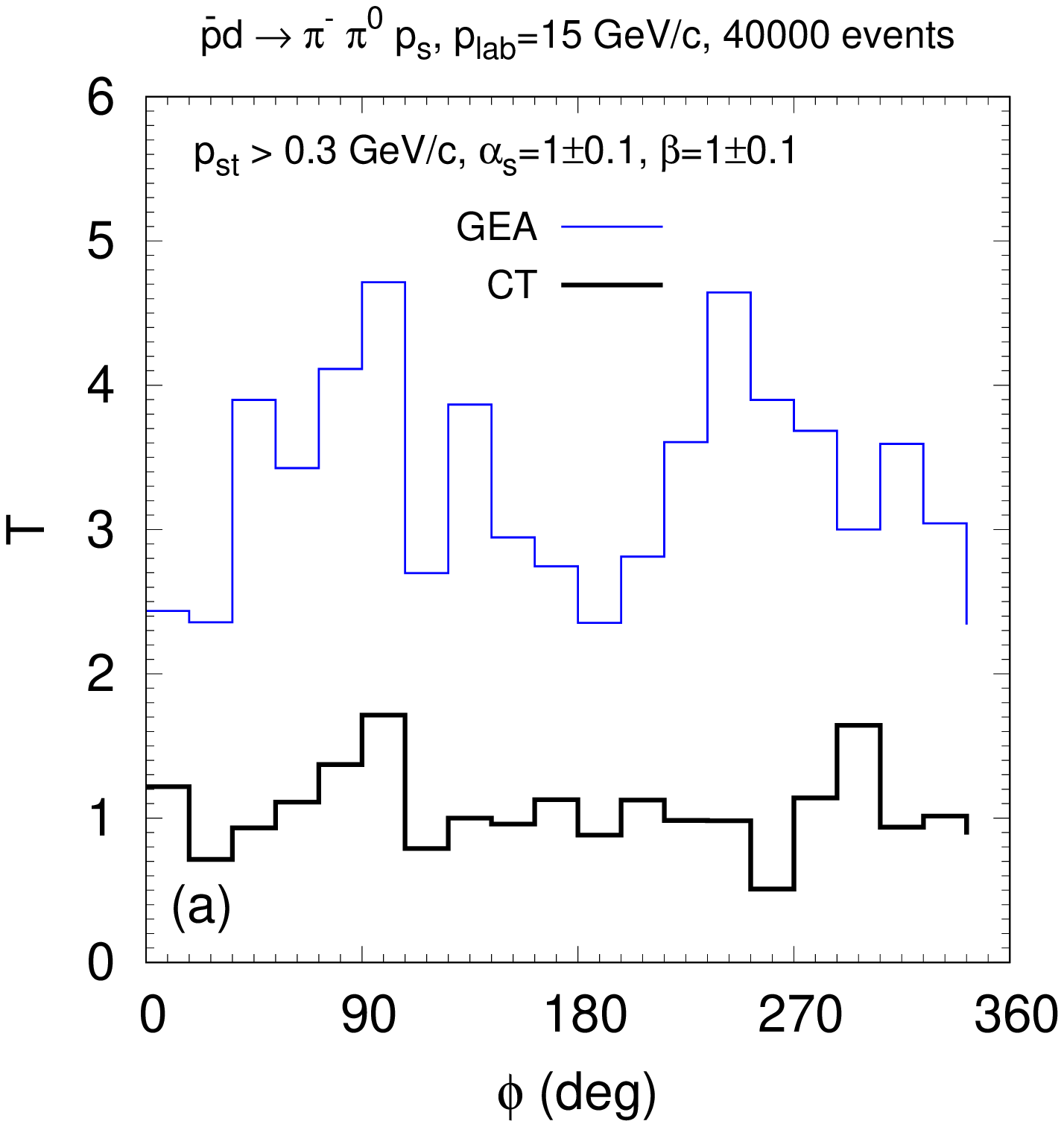}
  \includegraphics[scale = 0.50]{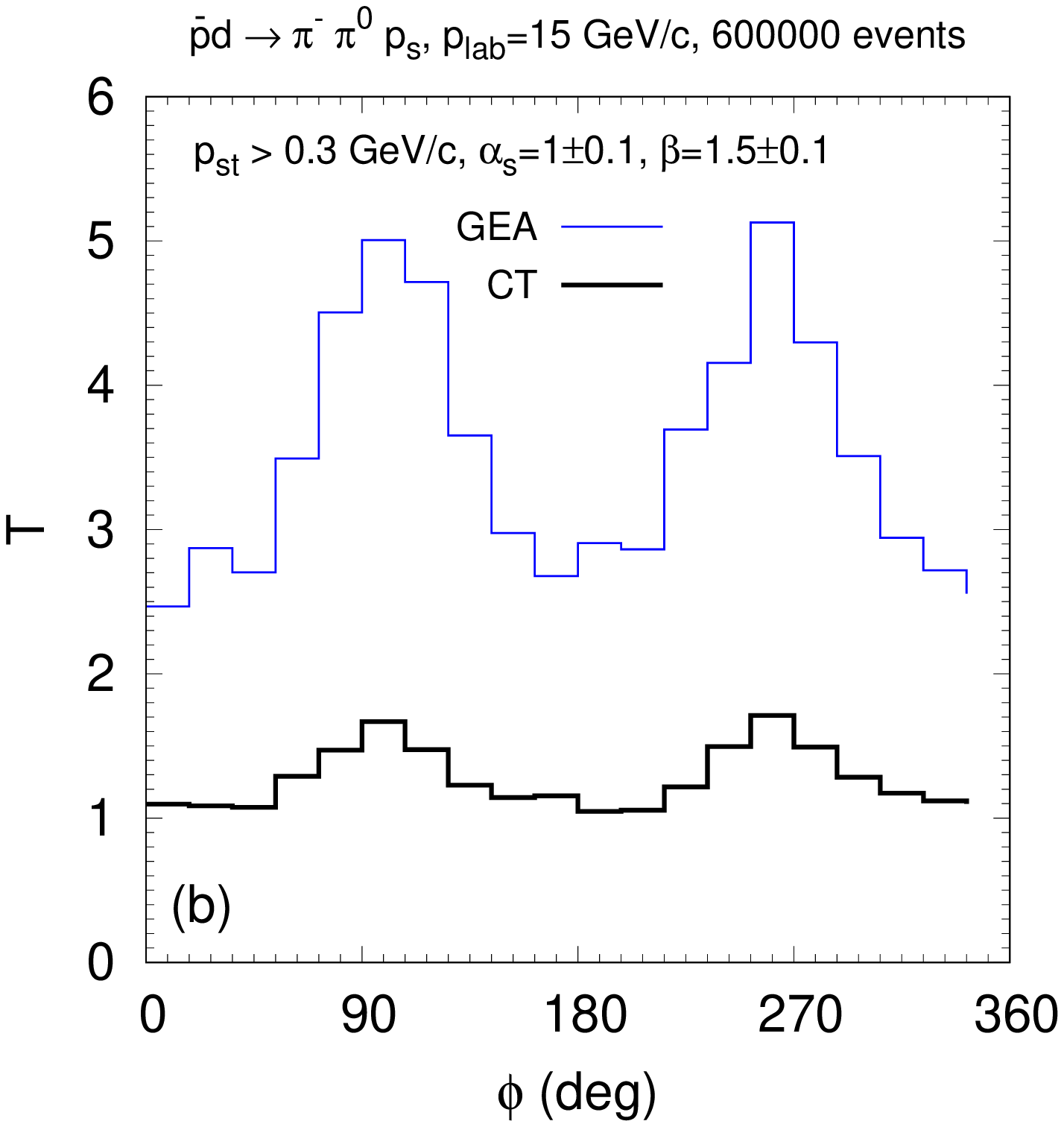}  
  \caption{\label{fig:Evdist_15gevc_phiDep_pst0.3gev}
    Transparency ratio from the MC simulation at the beam momentum of 15 GeV/c, $\alpha_s=1\pm0.1$ 
    as a function of the relative azimuthal angle between $\pi^-$
    and spectator proton for the transverse momentum of the spectator proton larger than 0.3 GeV/c.
    Panels (a) and (b) correspond to the kinematics with $\beta=1\pm0.1$ and $\beta=1.5\pm0.1$.
    Line notations are the same as in Fig.~\ref{fig:Evdist_15gevc}.}
\end{figure}
The study of the azimuthal distributions at large transverse momenta requires larger statistics
which makes the kinematics with $\beta=1.5\pm0.1$ preferable, as shown in Fig.~\ref{fig:Evdist_15gevc_phiDep_pst0.3gev}.
However, even for $\alpha_s=1\pm0.1$ we observe the same pattern as we found in the direct calculation of the 
distributions, see Fig.~\ref{fig:T_phiDep_15gevc} for $p_{st}=0.3$ GeV/c and 0.4 GeV/c.
It is quite clear that in calculations with CT the peaks at $90\degree$ and $270\degree$ are much less pronounced
since pion rescattering amplitude is suppressed.

\section{Summary and outlook}
\label{summary}

Based on the GEA method, we have developed the model for the exclusive channel $\bar p d \to \pi^- \pi^0 p$
at $p_{\rm lab} \sim 10$ GeV/c for large momentum transfer in the elementary process $\bar p n \to \pi^- \pi^0$.
The IA amplitude and the three amplitudes with elastic rescattering of either $\bar p$, $\pi^-$ or $\pi^0$
on the spectator proton have been included coherently. The CT effects are taken into account in the rescattering
amplitudes. The calculations of the four-fold differential cross section and of the transparency ratio have been
performed at  $p_{\rm lab}=5, 10$ and 15 GeV/c for small longitudinal momentum of spectator proton ($\alpha_s=1$)
and transverse momentum $p_{st} < 0.5$ GeV/c. The main results of our calculations can be summarized as follows:

\begin{itemize}
  
\item In a qualitative agreement with previous studies of the channel
$p d \to p p n$ with spectator neutron  \cite{Frankfurt:1996uz}, it is shown that the transparency ratio
is below one for small transverse momenta of the spectator ($p_{st} \ltsim 0.3$ GeV/c) indicating {\it absorption region}
and grows substantially above one for larger transverse momenta indicating {\it rescattering region}.

\item The transparency ratio as a function of the relative azimuthal angle $\phi$ between $\pi^-$ and spectator is studied.
The interference of the pion rescattering amplitude with the IA amplitude results in the transparency ratio having minima 
for small $p_{st}$ in the out-of-plane kinematics, i.e. for $\phi=90\degree$ and $270\degree$. On the other hand, at large $p_{st}$,
the pion rescattering amplitude squared produces the maxima in the transparency ratio at $\phi=90\degree$ and $270\degree$.
Studies of these observables would allow rigorous tests of GEA at the low end of the energy interval we considered.

\item CT leads to a factor of 2-3 increase (decrease) of transparency in the absorption (rescattering) region,
more pronounced at higher $p_{\rm lab}$.

\item The MC simulations indicate that the CT effects can be visible already with quite modest statistics
  of $\sim$ several 10000 events in the kinematical region of interest which could be collected during
  a pretty short run of PANDA with the deuteron target.

\end{itemize}

We have also checked that very similar behaviour appears in the channel $\bar p d \to \pi^- \pi^+ n$ with spectator
neutron. In some sense, theoretical predictions for that channel would be even more robust because the elementary process
$\bar p p \to \pi^- \pi^+$ is constrained by experiment. However, the detection of slow neutrons seems to be problematic at PANDA.
Thus, we decided to present only calculations for the channel with spectator proton in this paper. This channel would be the
simplest one and its study seems to be feasible at the beginning of PANDA operation. There are, of course, other opportunities to study 
CT, for example, in the channels $\bar p d \to K^- K^0 p$ and $\bar p d \to \pi^- \gamma p$. (In principle, there are much more possible channels,
but for these two the elastic scattering cross sections on the proton are known.)  In the latter channel, the photon transparency
(cf. \cite{Larionov:2016mim,Strikman:2017clu}) can also be studied.

The extension for the nuclear targets heavier than deuteron is a natural next step in theoretical studies of CT 
that is expected to significantly reduce absorption in semiexclusive channels with heavier targets, like $A(\bar p, \pi^- \pi^0)(A-1)^*$
and similar with other two-meson or meson-photon final states. Yet another interesting opportunity is the nuclear transparency
in the $A(\bar p, \bar p p)(A-1)^*$ quasi-elastic channel with large momentum transfer in the $\bar p p \to \bar p p$ scattering.
Since in the latter process, in contrast to the $pp \to pp$ scattering, the quark exchange is impossible, the PLC may not be formed
and then CT will not present \footnote{Remember that the exclusive large angle reactions for which a quark exchange in not allowed are strongly suppressed
  as compared to the ones for which the quark exchange is allowed \cite{White:1994tj}.}. Testing this expectation at PANDA would certainly be important
for our present understanding of the CT phenomenon.  

\begin{acknowledgments}
  We thank A.~Gillitzer, J.~Haidenbauer, and J.~Ritman for illuminating discussions.
  The research of M.S. was supported by the U.S. Department of Energy,
  Office of Science, Office of Nuclear Physics, under Award No. DE-FG02-93ER40771.
  The most of numerical calculations in the present work have been performed using the computational resources
  of the Frankfurt Center for Scientific Computing (FUCHS-CSC).
\end{acknowledgments}

\bibliography{pbard2pi}

\newpage

\appendix

\begin{appendices}

\section{Elementary amplitudes}
\label{ElemAmpl}  

\begin{subappendices}

\renewcommand{\thesubsection}{A\arabic{subsection}}

\subsection{$\bar p n \to \pi^- \pi^0$}
\label{M_ann}

\begin{figure}
  \includegraphics[scale = 0.60]{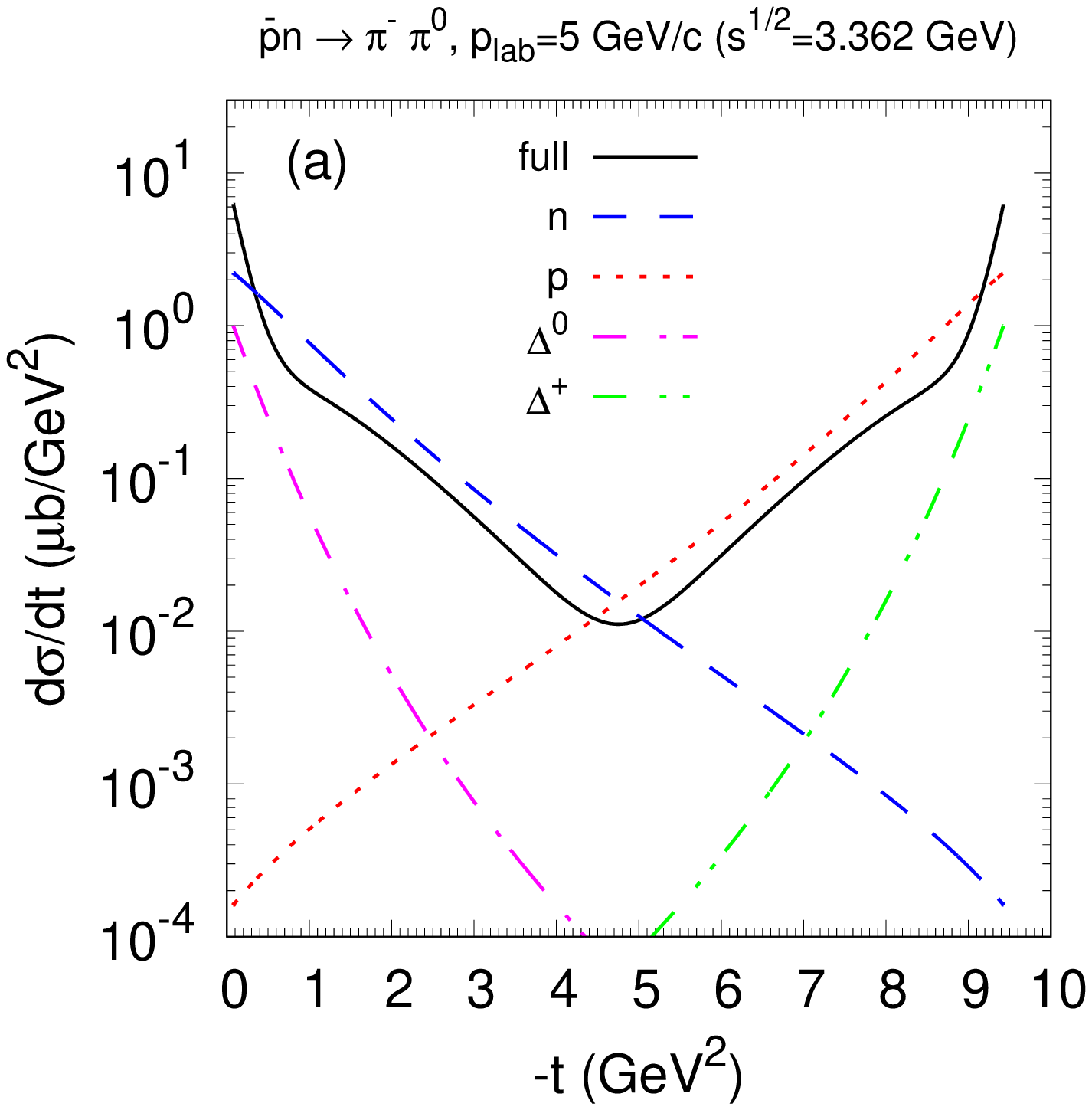}
  \includegraphics[scale = 0.60]{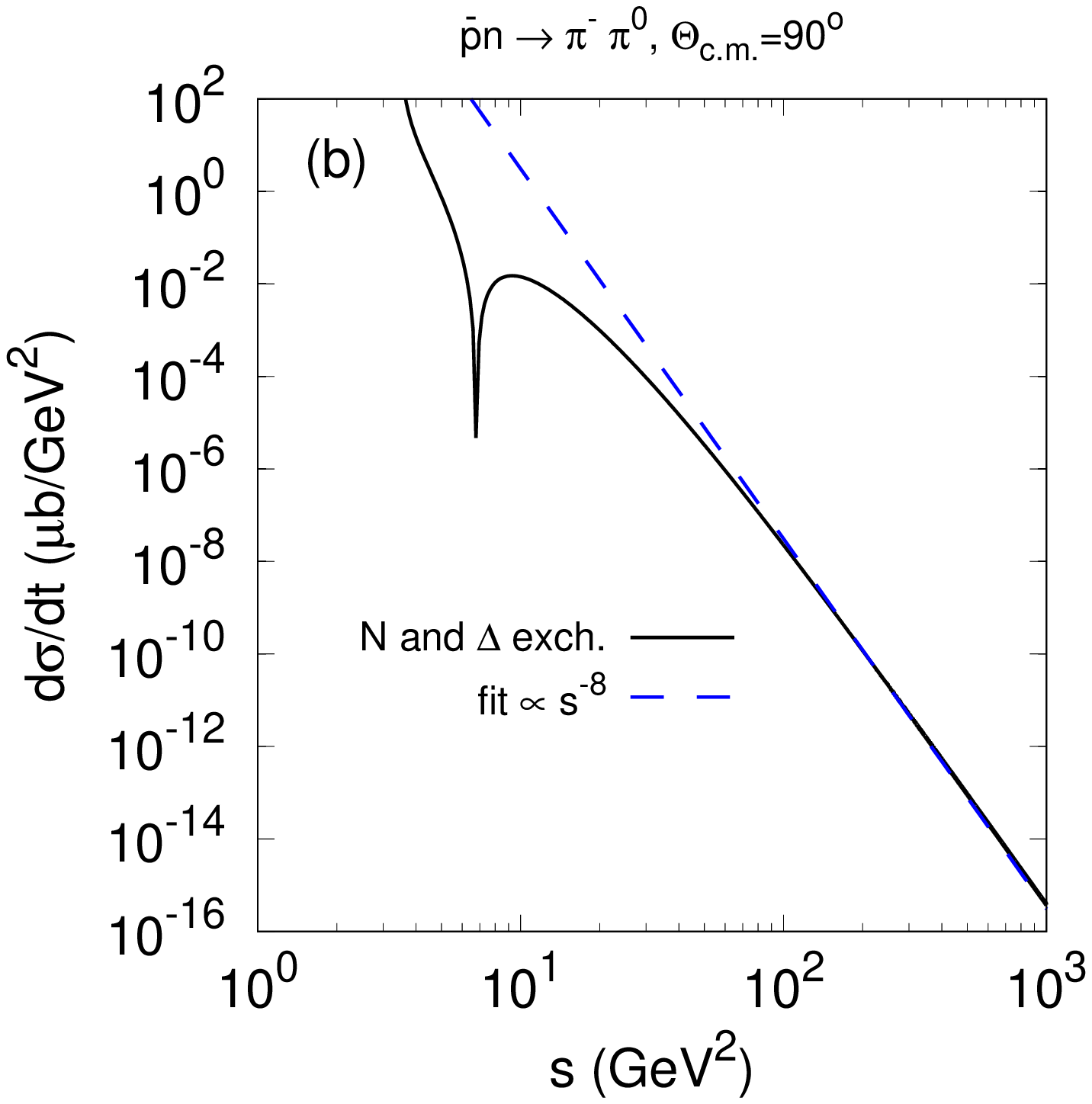}
  \caption{\label{fig:pbarn2pi-pi0} Differential cross section $\bar p n \to \pi^- \pi^0$ as a function of $-t$ at
    $p_{\rm lab}=5$ GeV/c (a) and as a function of $s$ at $\Theta_{\rm c.m.}=90\degree$ (b).
    In panel (a), the solid, dashed, dotted, dash-dotted and dash-double-dotted lines
    correspond to full cross section and to the partial contributions of the neutron, proton,
    $\Delta^0$ and $\Delta^+$ exchanges.
    In panel (b) the solid line shows the calculated full cross section while the dashed line --
    the power law fit of Eq.(\ref{counting}) with  $n=8$.}
\end{figure}

For the $\bar N N \to \pi \pi$ annihilation amplitude we apply the nucleon and $\Delta$ exchange model in a version
described in Appendix C of ref. \cite{Larionov:2018lpk}. The powers of the  vertex form factors are chosen from the
condition of the $s \to \infty,~t/s=\mbox{const}$ asymptotic scaling law \cite{Brodsky:1973kr,Matveev:1973ra}
\begin{equation}
   \frac{d\sigma}{dt} = \frac{f(t/s)}{s^n}~,~~~n=\sum n_i - 2~,     \label{counting}
\end{equation}
with $n_i$ being the number of quarks in each incoming and outgoing hadron. This leads to the
$\pi N N$ and $\pi N \Delta$ vertex form factors having powers of 2 and 5/2, respectively. The cutoff parameters
$\Lambda_{\pi NN}=2.0$ GeV and $\Lambda_{\pi N\Delta}=1.8$ GeV are chosen to reproduce the shape of the $t$-dependence
of the differential cross section $\bar p p \to \pi^- \pi^+$ at $p_{\rm lab}=5$ GeV/c (see Fig.~14 in ref. \cite{Larionov:2018lpk}).
The absolute value of that cross section at $\Theta_{\rm c.m.}=90\degree$ is accounted for by multiplying the invariant amplitude by the 
factor $\sqrt{\Omega}$ with $\Omega=0.008$ motivated by significant absorption in the incoming $\bar N N$ channel.
It is clear that such a description of the elementary $\bar N N \to \pi \pi$ amplitude is pretty simple.
However, we believe that it is good enough for our purposes to address reactions at $p_{\rm lab} \sim 5-15$ GeV/c.
Fig. \ref{fig:pbarn2pi-pi0} shows the $t$-dependence of the $\bar p n \to \pi^- \pi^0$
differential cross section $d\sigma/dt$ at $p_{\rm lab}=5$ GeV/c (a) and the $s$-dependence of the same cross section
at $\Theta_{\rm c.m.}=90\degree$ (b). The nucleon exchange contributions dominate at $\Theta_{\rm c.m.}=90\degree$ while
the $\Delta$ exchanges are important at forward and backward scattering angles.
The local minimum at $s=6.77$ GeV$^2$ ($p_{\rm lab}=2.5$ GeV/c) is due to destructive interference of $n$ and $p$
exchanges. In the case if only neutron exchange is possible ($\bar p p \to \pi^- \pi^+$) or the neutron and proton
exchanges interfere constructively ($\bar p p \to \pi^0 \pi^0$) the $s$-dependence is smooth.

\subsection{$\bar p p \to \bar p p$}
\label{M_pbarp}

The antiproton-proton elastic scattering amplitude at high energies and forward scattering angles can be conveniently
parameterized by the expression
\begin{equation}
    M_{\bar p p}(t)= 2 i p_{\rm lab} m_N \sigma_{\bar p p}^{\rm tot}
     (1-i\rho_{\bar p p}) \mbox{e}^{B_{\bar p p}t/2}~,           \label{M_barpp}
\end{equation}
where the $p_{\rm lab}$-dependent parameterizations $\sigma_{\bar p p}^{\rm tot}$ and $B_{\bar p p}$ are described
in ref. \cite{Larionov:2016xeb}. While the total cross section and the slope of momentum transfer dependence are
well constrained by experiment, the data on $\rho_{\bar p p}=\mbox{Re}M_{\bar p p}(0)/\mbox{Im}M_{\bar p p}(0)$
at $p_{\rm lab}=5-15$ GeV/c are quite scarce and thus we rely here on the extrapolation of the Regge-Gribov fit \cite{Patrignani:2016xqp}.
towards low beam momenta.
Fortunately, the sensitivity of our results to $\rho_{\bar p p}$ is quite weak.

\subsection{$\pi N \to \pi N$ elastic}
\label{M_piN}

Elastic scattering of charged pions on protons is thoroughly studied and the elastic amplitude can be thus
parameterized in a usual way:
\begin{equation}
    M_{\pi^\pm p}(t)= 2 i p_{\rm lab} m_N \sigma_{\pi^\pm p}^{\rm tot}
     (1-i\rho_{\pi^\pm p}) \mbox{e}^{B_{\pi^\pm p}t/2}~.           \label{M_pipmp}
\end{equation}
The $p_{\rm lab}$-dependent total $\pi^\pm p$ cross sections are well described by the
CERN-HERA fit \cite{Montanet:1994xu}.
The ratios $\rho_{\pi^\pm p}=\mbox{Re}M_{\pi^\pm p}(0)/\mbox{Im}M_{\pi^\pm p}(0)$ are taken from the Regge-Gribov fit
\cite{Patrignani:2016xqp}. The parameterizations of the slope parameters $B_{\pi^\pm p}$ at $|t|=0.2$ GeV$^2$
are provided in ref. \cite{Burq:1982ja}. The amplitude of the $\pi^0 p$ elastic scattering can be calculated from the
isospin relation:
\begin{equation}
      M_{\pi^0 p}(t)=\frac{1}{2}(M_{\pi^- p}(t)+M_{\pi^+ p}(t))~.            \label{M_pi0p}
\end{equation}

\end{subappendices}

\section{The differential cross section in the LC variables}  
\label{LCderiv}

The purpose of this Appendix is to derive Eq.(\ref{dsig/dalpha}). Let us consider the frame where
the four-momentum of the $\bar p + d$ system is ${\cal P} = p_{\bar p} + p_d = ({\cal P}^0,\bvec{0},P)$ with $P \to +\infty$.
In that frame the four-momenta of the pions are
$k_i=(\omega_i,\bvec{k}_{it},\tilde\alpha_i P),~i=1,2$ and the four-momentum of the spectator is
$p_s=(E_s,\bvec{p}_{st},\tilde\alpha_s P)$. The particle energies can be written as
\begin{eqnarray}
  \omega_i &=& \tilde\alpha_i P + \frac{m_{it}^2}{2\tilde\alpha_i P},~~~m_{it}^2=m_\pi^2+\bvec{k}_{it}^2~, \\
       E_s &=& \tilde\alpha_s P + \frac{m_{st}^2}{2\tilde\alpha_s P},~~~m_{st}^2=m_N^2+\bvec{p}_{st}^2~, \\
  {\cal P}^0 &=& P + \frac{{\cal P}^2}{2P}~.
\end{eqnarray}       
After simple algebra the invariant phase space volume element (\ref{dPhi_3}) can be rewritten in terms of the LC variables $\tilde\alpha_i$,
$i=1,2,s$ and the transverse momenta as follows:
\begin{eqnarray}
    d\Phi_3 &=&
     = 2\delta({\cal P}^2 - \frac{m_{1t}^2}{\tilde\alpha_1} - \frac{m_{2t}^2}{\tilde\alpha_2} - \frac{m_{st}^2}{\tilde\alpha_s})    \nonumber \\ 
     &&\times  \delta^{(2)}(\bvec{k}_{1t}+\bvec{k}_{2t}+\bvec{p}_{st}) \delta(1-\tilde\alpha_1-\tilde\alpha_2-\tilde\alpha_s)
    \frac{d^2k_{1t}d\tilde\alpha_1}{(2\pi)^32\tilde\alpha_1}  \frac{d^2k_{2t}d\tilde\alpha_2}{(2\pi)^32\tilde\alpha_2}
    \frac{d^2p_{st}d\tilde\alpha_s}{(2\pi)^32\tilde\alpha_s}~.     \label{dPhi_3_LC}
\end{eqnarray}
After successive integrations over $d^2k_{2t}d\tilde\alpha_2$ and $dk_{1t}$ the phase space volume (\ref{dPhi_3_LC}) becomes:
\begin{equation}
  d\Phi_3 = \frac{2}{|\partial{\cal F}/\partial k_{1t}|} \frac{k_{1t}d\phi_1d\tilde\alpha_1}{(2\pi)^32\tilde\alpha_1}
           \frac{1}{(2\pi)^32\tilde\alpha_2} \frac{p_{st}dp_{st}d\phi_s d\tilde\alpha_s}{(2\pi)^32\tilde\alpha_s}~,    \label{dPhi_3_integr}
\end{equation}
where
\begin{equation}
  {\cal F} = {\cal P}^2 - \frac{m_{1t}^2}{\tilde\alpha_1} - \frac{m_{2t}^2}{\tilde\alpha_2} - \frac{m_{st}^2}{\tilde\alpha_s}    \label{calF}
\end{equation}
and, therefore,
\begin{equation}
   \frac{\partial{\cal F}}{\partial k_{1t}} = -\frac{\partial}{\partial k_{1t}}\left(\frac{m_{1t}^2}{\tilde\alpha_1} + \frac{m_{2t}^2}{\tilde\alpha_2}\right)
   = - 2(k_{1t}/\tilde\alpha_1+(k_{1t}+p_{st}\cos\phi)/\tilde\alpha_2)~.   \label{partialcalF}
\end{equation}
For the non-polarized particles the differential cross section is invariant with respect to rotations about beam axis. Thus, we can integrate the cross section
over $d\phi_s$ keeping the relative azimuthal angle $\phi$ of Eq.(\ref{phi}) fixed. This leads to the following expression for the four-differential cross section:
\begin{equation}
  \tilde\alpha_s \tilde\alpha_1 \frac{d^4\sigma}{d\tilde\alpha_s\, d\tilde\alpha_1\, d\phi\, p_{s t} dp_{s t}}
  = \frac{\overline{|M|^2} k_{1t}}%
         {16(2\pi)^4 p_{\rm lab} m_d  |\partial{\cal F}/\partial k_{1t}| \tilde\alpha_2}~.                   \label{dsig/dtildealpha}
\end{equation}
By using the relations between the LC variables
\begin{eqnarray}
  \left|\frac{d\tilde\alpha_1}{\tilde\alpha_1}\right| &=& \left|\frac{d\beta}{\beta}\right| = \left|\frac{ dk_1^z}{\omega_1}\right|~,   \\
  \left|\frac{d\tilde\alpha_s}{\tilde\alpha_s}\right| &=& \left|\frac{d\alpha_s}{\alpha_s}\right| = \left|\frac{dp_s^z}{E_s}\right|
\end{eqnarray}
we finally obtain Eq.(\ref{dsig/dalpha}) where $\kappa_t=|\partial{\cal F}/\partial k_{1t}| \tilde\alpha_2$.

\end{appendices}

\end{document}